\documentclass[a4paper,oneside]{article}

\usepackage{microtype}
\usepackage[sc]{mathpazo}
\usepackage[T1]{fontenc}
\usepackage[DIV=11,BCOR=0cm,headinclude=true]{typearea}

\usepackage{amsfonts}
\usepackage{amsmath}
\usepackage{amssymb}
\usepackage{amsthm}
\usepackage{graphicx}
\usepackage{float}
\usepackage[config,textfont=sl]{caption,subfig}
\usepackage[british]{babel}
\usepackage{bm}

\usepackage[maxbibnames=100,autocite=superscript,style=custom-nature,sorting=none,defernumbers=true]{biblatex}
\usepackage{authblk}

\newcommand{\bv}[1]{\boldsymbol{#1}}
\newcommand{\bra}[1]{\langle{#1}|}
\newcommand{\ket}[1]{|{#1}\rangle}

\newcommand{\tensor}[1]{\bar{\bar{#1}}}

\DeclareMathOperator*{\real}{Re}
\DeclareMathOperator*{\imag}{Im}

\author[1,$\dagger$]{Parry Y.\ Chen}
\author[1]{Yonatan Sivan}
\affil[1]{School of Electrical and Computer Engineering, Ben-Gurion University, Israel}
\affil[$\dagger$]{\it{parryyu@post.bgu.ac.il}}

\def\XXint#1#2#3{{\setbox0=\hbox{$#1{#2#3}{\int}$}
     \vcenter{\hbox{$#2#3$}}\kern-.5\wd0}}

\title{Resolving the Gibbs phenomenon via a discontinuous basis in a mode solver for open optical systems}
\date{\today}

\begin{document}
\maketitle

\begin{abstract}
Partial differential equations are frequently solved using a global basis, such as the Fourier series, due to excellent convergence. However, convergence becomes impaired when discontinuities are present due to the Gibbs phenomenon, negatively impacting simulation speed and possibly generating spurious solutions. We resolve this by supplementing the smooth global basis with an inherently discontinuous basis, incorporating knowledge of the location of the discontinuities. The solution's discontinuities are reproduced with exponential convergence, expediting simulations. The highly constrained discontinuous basis also eliminates the freedom to generate spurious solutions. We employ the combined smooth and discontinuous bases to construct a solver for the modes of a resonator in an open electromagnetic system. These modes can then expand any scattering problem for any source configuration or incidence condition without further numerics, enabling ready access and physical insight into the spatial variation of Green's tensor. Solving for the modes is the most numerically intensive and difficult step of modal expansion methods, so our mode solver overcomes the last major impediment to the use of modal expansion for open systems.
\end{abstract}

\section{Introduction}
Partial differential equations (PDEs) are ubiquitous across all of engineering and physics, and their numerical solution is required for the majority of applications. A common feature of many physical systems is the presence of sharp discontinuities in material parameters corresponding to an interface. Solutions are piecewise smooth, but the presence of discontinuities pose great difficulties for many numerical methods, restricting the utility of otherwise excellent methods. In particular, global basis expansion methods, such as Fourier expansion methods,\autocite{kreyszig2011advanced,stein2003fourier} are often fast, accurate, and reliable. These traits are underpinned by their exponential convergence, but this much coveted property is crippled in the presence of discontinuities due to the Gibbs phenomenon.\autocite{strang2007computational}

Many successful solutions to the Gibbs phenomenon have been developed, given knowledge of the locations of the discontinuities. One such technique is reconstruction by Gegenbauer polynomials,\autocite{gottlieb1992gibbs,gottlieb2011review,gottlieb1997gibbs} and other reprojection methods.\autocite{driscoll2001pade,shizgal2003resolution,gelb2007reconstruction,adcock2012stable} Other techniques require knowledge of both the location and size of the discontinuity.\autocite{eckhoff1998high,wanguemert2007removal} Remarkably, these techniques can rapidly recover the true solutions from Fourier series tainted by Gibbs phenomena. Despite their success in 1D, reconstructions in higher dimensions are more problematic, requiring line-by-line treatment or tensor products,\autocite{min2006fourier,adcock2012stable} neither of which are easily adapted to curved or complex interfaces. Reconstruction techniques also do not necessarily accelerate the convergence if other high-frequency features are present, such as exponential decay. Such behavior is characteristic of bound modes and systems where material parameters change sign across an interface. Since reconstruction techniques rely on post-processing of the fields, other values such as eigenvalues of modes remain impacted by the Gibbs phenomenon. A second post-processing step would be required to deduce more accurate eigenvalues from the reconstructed eigenmodes.\autocite{wanguemert2007removal} Lastly, representing products of discontinuous functions can lead to utter failure of naive methods, requiring the use of complex factorization rules.\autocite{li1996use,neviere2002light}

In this paper, we resolve the Gibbs phenomenon by introducing an inherently discontinuous basis, to be used alongside the regular basis, which is usually smooth. Since the former is responsible for representing all the discontinuities, the convergence properties of the smooth basis are recovered. The discontinuous basis incorporates knowledge of the location of discontinuities, but no knowledge of the size of the discontinuities is necessary. Overall, a smaller basis is needed, smaller by several orders of magnitude compared to entirely smooth basis sets, even if only a few digits of accuracy are desired.\autocite{doost2014resonant,lobanov2019resonant} This minimizes the size of any linear system of equations to be solved, expediting numerical solution. The discontinuous basis is easily defined and readily obtained given knowledge of the location of the discontinuity. Curved interfaces are easily treated in two or three dimensional problems. Both the smooth and discontinuous basis sets are used to expand the solution from the outset, with no post-processing necessary. Perhaps of greater importance is that the tightly constrained discontinuous basis suppresses numerical noise, eliminating numerical artifacts and spurious solutions. Especially for the eigenmode problem, this yields a clean set of modes, without any need for manual identification of spurious modes.

We implement this basis to find the modes of an open electromagnetic system. We treat eigenvalue problems, which remain problematic for reconstruction methods. We treat an open system since global bases are advantageous for infinite domains, where the far-field response is efficiently represented by the multipole basis. Open systems were also difficult to treat using modal methods, though significant research progress has been achieved.\autocite{lalanne2018light,lalanne2019quasinormal,sehmi2020applying,ge2014quasinormal} We subsequently use our modes to simulate a scattering problem. Our inherently discontinuous modes can also be applied to finite closed systems or periodic systems, or to treat the scattering problem directly.

\begin{figure}[!t]
\begin{center}
\includegraphics[width=40mm]{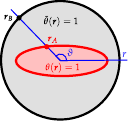}
\caption{Modes of an embedding geometry, such as a circle, are available analytically. These are used as a basis to solve for the modes of an enclosed target geometry, and ellipse in this case. To describe the target and embedding geometries for future use in Section \ref{sec:genome}, two functions are introduced. The Heaviside type function $\tilde{\theta}(\bv{r})$ is defined to be $1$ for the interior of the circle and zero everywhere else, and similarly $\theta(\bv{r})$ is defined to be $1$ in the interior of the ellipse in this example. The remaining quantities are used in Section \ref{sec:longitudinal}, with polar coordinates $(r,\vartheta)$. The vectors $\bv{r}_A$ and $\bv{r}_B$ are the locus of points along the boundaries of the target and basis geometries, respectively. }
\label{fig:embedding}
\end{center}
\end{figure}

We shall solve for a set of target modes $\{\bv{x}_m(\bv{r})\}$ by expanding each using a set of basis functions $\{\tilde{\bv{x}}_\mu(\bv{r})\}$,
\begin{equation}
\bv{x}_m(\bv{r}) = \sum_\mu c_{\mu,m} \tilde{\bv{x}}_\mu(\bv{r}),
\label{eq:expansiongen}
\end{equation}
The set $\{\tilde{\bv{x}}_\mu(\bv{r})\}$ is global, defined to be non-zero across the entire domain. This contrasts with basis functions for finite element methods for example,\autocite{jin2014finite} which are localized to each mesh element. The plane wave basis is an example of a frequently used global basis, leading to Fourier series solutions of PDEs,\autocite{neviere2002light} with another possibility being Wannier functions.\autocite{busch2003wannier} To find the coefficients $c_{\mu,m}$, the series \eqref{eq:expansiongen} is inserted into the defining PDE for the target modes, and by projecting back onto the basis, overlap integrals are obtained,
\begin{equation}
V_{\nu\mu} = \bra{\tilde{\bv{x}}_\nu}\hat{V}\ket{\tilde{\bv{x}}_\mu},
\label{eq:overlapbraket}
\end{equation}
where $\hat{V}$ represents the PDE, the geometry, and material parameters. The elements $V_{\nu\mu}$ fill a dense linear system of equations, solved using a numerical linear algebra package. Their dense nature is often offset by their small size due to rapid convergence.

In this manuscript, we treat the modes of finite inclusions in an infinite domain. Thus, Fourier series are difficult to use as the basis functions $\{\tilde{\bv{x}}_\mu(\bv{r})\}$, since a continuum would be necessary, transforming the sum in \eqref{eq:expansiongen} to an integral.\autocite{kreyszig2011advanced,stein2003fourier} We choose a more efficient basis, defining the modes of a simpler open system as the basis functions $\{\tilde{\bv{x}}_\mu(\bv{r})\}$, whose modes are easily obtained.\autocite{doost2013resonant,chen2020efficient}. A 2D example is given in Figure \ref{fig:embedding}, where the eigenmodes of an elliptical inclusion in an infinite background, $\{\bv{x}_m(\bv{r})\}$, are expanded using the eigenmodes of a circle in an infinite background, $\{\tilde{\bv{x}}_\mu(\bv{r})\}$. The latter modes still form a discrete set and are available analytically.\autocite{chen2017robust,reddy2017revisiting} Equivalently in 1D and 3D, the modes of a slab and sphere are available analytically,\autocite{forestiere2016material,farhi2016electromagnetic,bergman1980theory} and these are the natural embedding geometries that can be used to generate the modes of more complex geometries in 1D and 3D. We shall refer to the circle as the embedding geometry, and to its modes as embedding modes or basis modes. Its interior encloses the target elliptical geometry, whose modes are the target modes. Despite the infinite domain, the overlap integrals \eqref{eq:overlapbraket} are evaluated only over the target geometry. Since the embedding modes $\{\tilde{\bv{x}}_\mu(\bv{r})\}$ are used to expand the target modes $\{\bv{x}_m(\bv{r})\}$, which themselves are subsequently used for expansion, we term the first process \emph{re-expansion}. The terminology \emph{resonant state expansion} has also been used by other authors.\autocite{doost2013resonant,doost2014resonant}

One of the benefits of first finding the modes $\{\bv{x}_m(\bv{r})\}$ of an inclusion is that any given scattering problem involving this inclusion then requires no further numerical computation to solve.\autocite{chen2019generalizing,bergman1980theory,muljarov2010brillouin, doost2013resonant, lalanne2018light, yan2018rigorous} The solution is simply given by a sum over the eigenmodes, each weighted by the spatial overlap between the eigenmode and the source, and the detuning between the actual physical parameters from the eigenvalues.\autocite{chen2019generalizing,lalanne2018light} These eigenmode expansion methods are ideally suited to repeated simulations with many different source configurations. They have received intensive research effort over recent years in the context of open electrodynamic systems, and many theoretical and practical challenges have been overcome.\autocite{lalanne2018light,lalanne2019quasinormal,sehmi2020applying} Yet modal expansions in open systems remain hindered by the lack of reliable mode solvers. We surmount this last major obstacle, creating a powerful and practical method for repeated simulations, paving the path towards widespread adoption of modal expansion methods.

\section{Maxwell's equation via generalized normal mode expansion}
\label{sec:genome}
We begin by providing the context in which we introduce our discontinuous basis. We aim to find the modes of Maxwell's equations in an open system that provides the solution for any arbitrary source $\bv{J}(\bv{r})$,
\begin{equation}
\nabla\times(\nabla\times\bv{E}) - k^2\epsilon(\bv{r})\bv{E} = ikZ_0\bv{J},
\label{eq:maxwell}
\end{equation}
where $k = \omega/c$ and $Z_0 = \sqrt{\mu_0/\epsilon_0}$ is the impedance of free space in SI units. We have assumed harmonic $e^{-i\omega t}$ time variation, and also assume non-magnetic media across the whole domain. The structure is defined by its permittivity profile $\epsilon(\bv{r})$, which we assume consists of a finite inclusion of uniform permittivity $\epsilon_i$ of arbitrary shape, resting in an infinite background of uniform permittivity $\epsilon_b$, such that
\begin{equation}
\epsilon(\bv{r}) = \epsilon_b + (\epsilon_i - \epsilon_b)\theta(\bv{r}),
\end{equation}
where $\theta(\bv{r})$ is a step function which is unity within the interior of the inclusion and zero elsewhere (see Figure \ref{fig:embedding}). We restrict attention here to piecewise uniform structures, since this is where a discontinuous basis has the greatest impact. For spatially non-uniform inclusions, the techniques of Refs \parencite{chen2020efficient,agranovich1999generalized} can be applied.

The modes we desire are eigenmodes of the Lippmann-Schwinger equation for electromagnetism, which can be obtained by manipulating \eqref{eq:maxwell} to yield
\begin{equation}
\nabla\times(\nabla\times\bv{E}) - k^2\epsilon_b\bv{E} = ikZ_0\bv{J} + k^2\theta(\bv{r})(\epsilon_i-\epsilon_b)\bv{E},
\label{eq:inhomowave}
\end{equation}
where the second term on the right hand side represents the displacement current. We apply the Green's function, which characterizes the response to a unit impulse, to transform the differential equation \eqref{eq:inhomowave} to integral form. Since the left hand side \eqref{eq:inhomowave} is now spatially invariant, the Green's function of uniform space, $\tensor{G}_0(|\bv{r}-\bv{r}'|)$,\autocite{novotny2012principles} may be applied to yield the desired Lippmann-Schwinger equation,
\begin{equation}
\bv{E}(\bv{r}) = \bv{E}_0(\bv{r}) + k^2 \int \tensor{G}_0 (|\bv{r} - \bv{r}'|) (\epsilon(\bv{r}') - \epsilon_b) \bv{E}(\bv{r}')\, d\bv{r}'.
\label{eq:lippsch}
\end{equation}
The term $\bv{E}_0(\bv{r})$ may either be specified or readily obtained from $\bv{J}(\bv{r})$ via
\begin{equation}
\bv{E}_0(\bv{r}) = i\omega\mu_0 \int \tensor{G}_0(|\bv{r} - \bv{r}'|) \bv{J}(\bv{r}')\, d\bv{r}',
\label{eq:E0}
\end{equation}
since it is the known radiation pattern of $\bv{J}(\bv{r})$ in a uniform background

In \eqref{eq:lippsch}, the unknown solution $\bv{E}(\bv{r})$ appears both inside and outside the integral, which can be solved projecting onto an appropriate set of normal modes, obtained by neglecting the source terms of \eqref{eq:inhomowave} or \eqref{eq:lippsch}. We choose to show the differential form,
\begin{equation}
\nabla\times(\nabla\times\bv{E}_m) - k^2\epsilon_b\bv{E}_m = \frac{1}{s_m}k^2\epsilon_b\theta(\bv{r})\bv{E}_m,
\label{eq:eigendiff}
\end{equation}
where $s_m$ is the $m$th eigenvalue,
\begin{equation}
\frac{1}{s_m} \equiv \frac{\epsilon_m - \epsilon_b}{\epsilon_b},
\label{eq:eigenvalue}
\end{equation}
and $\epsilon_m$ is the eigenpermittivity of the mode.\autocite{bergman1980theory, agranovich1999generalized, chen2019generalizing, sandu2012eigenmode, ge2010steady, forestiere2016material} This provides the modal expansion solution of Maxwell's equation suitable for open geometries,\autocite{bergman1980theory, agranovich1999generalized, chen2019generalizing} which after some manipulation is
\begin{equation}
\bv{E}(\bv{r}) = \bv{E}_0(\bv{r}) + \frac{i}{\omega\epsilon_0} \sum_m \bv{E}_m(\bv{r})\frac{\epsilon_i-\epsilon_b}{(\epsilon_m-\epsilon_i)(\epsilon_m-\epsilon_b)} \int \bv{E}^\dagger_m(\bv{r}') \bv{J}(\bv{r}')\, d\bv{r}',
\label{eq:genome}
\end{equation}
where the adjoint mode is simply the transpose of the direct mode, 
\begin{equation}
\bv{E}^\dagger_m(\bv{r}) = \bv{E}^\intercal_m(\bv{r}).
\end{equation}
Thus, the solution for any source configuration can be obtained almost immediately once the modes \eqref{eq:eigendiff} are available by evaluating the dot product $\bv{E}^\dagger_m(\bv{r}) \bv{J}(\bv{r})$. An alternative form of \eqref{eq:genome} is available if $\bv{E}_0(\bv{r})$ is specified instead of $\bv{J}(\bv{r})$.\autocite{bergman1980theory, chen2019generalizing}

\section{Solving for modes by re-expansion}
\label{sec:perturbation}
The bulk of the computational effort in any modal expansion method is typically devoted to finding the modes. We use the expansion
\begin{equation}
\bv{E}_m(\bv{r}) = \sum_\mu c_{\mu,m} \tilde{\bv{E}}_\mu(\bv{r}),
\label{eq:expansion}
\end{equation}
where $\bv{E}_m(\bv{r})$ is a target mode defined by \eqref{eq:eigendiff}, and $\tilde{\bv{E}}_\mu(\bv{r})$ are the embedding modes associated with a simpler open system,
\begin{equation}
\nabla\times(\nabla\times\tilde{\bv{E}}_\mu) - k^2\epsilon_b\tilde{\bv{E}}_\mu = \frac{1}{\tilde{s}_\mu}k^2\epsilon_b\tilde{\theta}(\bv{r})\tilde{\bv{E}}_\mu.
\label{eq:eigenb}
\end{equation}
The function $\tilde{\theta}(\bv{r})$ defines the interior of another geometry, and like $\theta(\bv{r})$, is a step function which is unity within a finite region and zero elsewhere (see Figure \ref{fig:embedding}). It is usually chosen to be a slab in 1D, a circle in 2D, or a sphere in 3D, so that its modal fields are available analytically.  For \eqref{eq:expansion} to be valid, the interior of $\tilde{\theta}(\bv{r})$ must enclose the interior of $\theta(\bv{r})$, as \eqref{eq:eigenb} provides a complete basis only within its interior. In the infinite background, \eqref{eq:eigenb} provides a complete basis only for outgoing fields, corresponding to a multipole expansion. Modes \eqref{eq:eigenb} of different eigenvalues $\tilde{s}_\mu$ obey an orthonormality relation\autocite{bergman1980theory, chen2019generalizing}
\begin{equation}
\int \tilde{\bv{E}}^\dagger_\nu(\bv{r}) \tilde{\theta}(\bv{r}) \tilde{\bv{E}}_\mu(\bv{r})\, d\bv{r} = \delta_{\nu\mu},
\label{eq:ortho}
\end{equation}
which is useful for projection.

The short derivation for obtaining the unknown coefficients $c_{\mu,m}$ is supplied in Ref.\ \parencite{chen2020efficient}. It involves inserting expansion \eqref{eq:expansion} into the target eigenvalue equation \eqref{eq:eigendiff} and projecting using orthonormality \eqref{eq:ortho} to obtain the linear eigenvalue problem
\begin{equation}
s_m c_{\nu,m} = \tilde{s}_\nu \sum_\mu V_{\nu\mu} c_{\mu,m},
\label{eq:perteig}
\end{equation}
where $\tilde{s}_\nu$ is the known eigenvalue of $\tilde{\bv{E}}_\nu(\bv{r})$, $s_m$ is the unknown eigenvalue of $\bv{E}_m(\bv{r})$, and $V_{\nu\mu}$ are the overlap integrals 
\begin{equation}
V_{\nu\mu} = \int \tilde{\bv{E}}^\dagger_\nu(\bv{r}) \theta(\bv{r}) \tilde{\bv{E}}_\mu(\bv{r})\, d\bv{r}
\label{eq:Vdef}
\end{equation}
among known basis modes over the target function $\theta(\bv{r})$. The matrix $V_{\nu\mu}$ is complex symmetric, so \eqref{eq:perteig} can be symmeterized by multiplying and dividing by $\tilde{s}$. Normalization also immediately follows.\autocite{chen2020efficient} The numerical implementation of \eqref{eq:perteig} first requires preparation of all the embedding modes, \eqref{eq:eigenb}, considered in Section \ref{sec:longitudinal}. 

\section{Discontinuous basis modes}
\label{sec:longitudinal}
The main aim of this section is to introduce a suitable embedding basis for the expansion \eqref{eq:expansion}. This entails the introduction of longitudinal modes, which we construct to be discontinuous. Before we embark on this task, we first consider the physical origins of field discontinuities in the context of Maxwell's equations, as this shall guide our construction.

Consider the divergence condition $\oint \bv{D} \cdot d\bv{S} = 0$ of Maxwell's equations, which is equal to zero for any mode since there are no external sources by definition. Undergraduate textbook arguments using a Gaussian pillbox lead to the condition that the normal component of $\bv{E}$ be discontinuous by the ratio of permittivities across an interface. For our purposes, it is more illuminating to consider the differential form
\begin{equation}
\nabla\cdot\bv{D} = \nabla\cdot(\epsilon(\bv{r})\bv{E}) = \epsilon(\bv{r})\nabla\cdot\bv{E} + \bv{E}\cdot\nabla\epsilon(\bv{r}) = 0.
\end{equation}
For piecewise uniform permittivity profiles, $\nabla\epsilon(\bv{r})$ is nonzero only where $\epsilon(\bv{r})$ is discontinuous, so we find
\begin{equation}
\nabla\cdot\bv{E} = -\frac{1}{\epsilon(\bv{r})}\bv{E}\cdot\nabla\epsilon(\bv{r}) = -\frac{\Delta\epsilon}{\epsilon(\bv{r})} \bv{E}\cdot\bv{\hat{n}}\, \delta(\bv{r}-\bv{r}_A),
\label{eq:divE}
\end{equation}
defining $\Delta\epsilon$ as the discontinuity in permittivity, and $\bv{r}_A$ as the locus of all points along the interface of $\theta(\bv{r})$, with $\bv{\hat{n}}$ being its unit normal vector.

The precise form of the divergence $\nabla\cdot\bv{E}$ in \eqref{eq:divE} is not as important as the following salient points. A discontinuity in $\bv{E}$ corresponds to a non-zero divergence, which is infinitesimally localized to the interface. Equivalently, the field is said to be non-transverse or have a longitudinal component. This conclusion is perhaps obvious, given that a layer of bound charges is known to be induced at interfaces, which causes the $\bv{E}$ field to be non-transverse here. Meanwhile, the divergence $\nabla\cdot\bv{E}$ is expected to be continuous along the interface $\bv{r}_A$, since the field $\bv{E}$ is everywhere continuous except at the interface. Corners and other sharp points along $\bv{r}_A$ are exceptions to this, since the field becomes singular, so the divergence $\nabla\cdot\bv{E}$ may be discontinuous or singular. From \eqref{eq:divE}, it is also clear that transverse basis modes are incapable of representing field discontinuities, and basis modes with a longitudinal component are necessary. Indeed, for piecewise uniform resonators, the sole purpose of longitudinal basis modes is to reproduce discontinuities, as the fields of the target mode are otherwise everywhere transverse.

To see whether the defining eigenvalue equation \eqref{eq:eigenb} possesses the necessary modes, we now analyze the types of modes that emerge. We recapitulate some important concepts from previous publications, but for further details these references should be consulted.\autocite{chen2020efficient, bergman1980theory, lobanov2019resonant} The first set of modes produced by \eqref{eq:eigenb} are the most familiar, corresponding to solutions with any eigenvalue except $\tilde{s}_\mu \neq -1$, or equivalently $\tilde{\epsilon}_\mu \neq 0$. These are transverse in the interior of $\tilde{\theta}(\bv{r})$, satisfying $\nabla\cdot\tilde{\bv{E}}_\mu = 0$, which can be shown by taking the divergence of \eqref{eq:eigenb}. Since there are a discrete infinity of such modes, they are able to represent the smooth near-field features and all the far-field features. For a planar, circular, or spherical embedding geometry $\tilde{\theta}(\bv{r})$, these modes can be found by solving a transcendental equation such as \eqref{eq:disprel}.\autocite{chen2017robust,reddy2017revisiting,forestiere2016material,farhi2016electromagnetic}

Another set of basis modes arises when the eigenvalue is $\tilde{s}_\mu = -1$ or $\tilde{\epsilon}_\mu = 0$, for which \eqref{eq:eigenb} collapses to $\nabla\times(\nabla\times\tilde{\bv{E}}_\mu) = 0$. Such modes are mathematically similar to zero frequency or static solutions. We focus on the subset arising from $\nabla\times\tilde{\bv{E}}_\mu = 0$, which feature longitudinal $\bv{E}$ fields capable of representing discontinuities.\autocite{chen2020efficient, bergman1980theory, lobanov2019resonant} These irrotational modes can be represented as a potential,
\begin{equation}
\tilde{\bv{E}}_\mu = \nabla\tilde{\phi}_\mu,
\label{eq:potendef}
\end{equation}
subject to the boundary conditions
\begin{equation}
\tilde{\phi}_\mu\biggr\rvert_{\partial B} = 0,
\label{eq:dirichlet}
\end{equation}
where $\partial B$ defines the interface between the interior and exterior of the embedding geometry $\tilde{\theta}(\bv{r})$. This boundary condition is a consequence of $\tilde{\epsilon}_\mu = 0$. From \eqref{eq:dirichlet}, it can also be shown that all non-trivial modes $\tilde{\phi}_\mu$ must have $\nabla\cdot\tilde{\bv{E}}_\mu \neq 0$ somewhere in the domain. Since these longitudinal modes originate from the same eigenmode equation \eqref{eq:eigenb} as the transverse modes, they complement and can be used alongside the transverse modes. This is simplified by the orthogonality relation \eqref{eq:ortho} between the transverse and longitudinal sets. Another useful property of these static-like modes is that they can be used with modes of any eigenvalue equation \eqref{eq:eigendiff}, regardless of the other parameters in \eqref{eq:eigendiff}.

A continuous infinity of zero eigenpermittivity modes exist, subject only to the conditions \eqref{eq:potendef} and \eqref{eq:dirichlet}. Great freedom exists in their construction to suit particular simulation needs. In previous publications,\autocite{chen2020efficient} a complete orthonormal set of longitudinal modes was defined by imposing the additional restriction that they be eigenmodes of the Laplace operator, 
\begin{equation}
\nabla^2\tilde{\phi}_\mu + \alpha_\mu^2\tilde{\phi}_\mu = 0.
\label{eq:FBmode}
\end{equation}
This results in smooth modes, and in the special case of circular and spherical domains, a Fourier-Bessel basis. Furthermore, their divergence is non-zero across the domain, $\nabla\cdot\tilde{\bv{E}}_\mu = \nabla^2\tilde{\phi}_\mu = -\alpha^2\tilde{\phi}_\mu \neq 0$. While \eqref{eq:FBmode} yields a complete set, in principle capable of representing any pattern of longitudinal fields, it suffers from severe Gibbs phenomenon when representing \eqref{eq:divE}, where fields are longitudinal only along the infinitesimal step-change region. In our case, use of \eqref{eq:FBmode} also produces many spurious target modes, which are longitudinal even within the uniform regions of the target geometry, where the field should be strictly transverse. Though perhaps easy to identify by eye via visual inspection of the modal fields, distinguishing and discarding such modes through automated methods can be difficult due to numerical noise.

At this point, we depart from the established literature and seek to define a more suitable, inherently discontinuous basis of longitudinal modes by exploiting the flexibility of \eqref{eq:potendef} and \eqref{eq:dirichlet}. To achieve this, we return to \eqref{eq:divE}, which stipulates that the field be longitudinal only along the interface of $\theta(\bv{r})$. While we know the location of the longitudinal component, we do not know its magnitude prior to solving for the modes. So the basis we desire should (a) be constrained to be longitudinal only along the interface, and (b) be able to capture the variation along the interface. This variation is also expected to be smooth, so long as the interface contains no corners.

For the 2D case, we assume that the interface is given in polar coordinates by $\bv{r}_A = (r, \vartheta) = (a(\vartheta), \vartheta)$, where $a(\vartheta)$ is a smooth function. These coordinates are displayed in Figure \ref{fig:embedding}. To satisfy the two requirements above, we construct a set of modes whose longitudinal components are all infinitesimally localized radially and vary as a Fourier series azimuthally,
\begin{equation}
\nabla\cdot\tilde{\bv{E}}_\lambda = \nabla^2\tilde{\phi}_\lambda = \frac{1}{2\pi r} \delta(r-a(\vartheta)) e^{i\lambda\vartheta},
\label{eq:Lmodecyl}
\end{equation}
where $\lambda$ is some integer angular order. The factor of $2\pi r$ ensures that the integral is unity, $\int \nabla^2\tilde{\phi}_\lambda\, d\bv{r} = 1$, though this is not strictly necessary as the modes require normalization according to \eqref{eq:ortho} before use. For the 3D case, the process is similar assuming that $\bv{r}_A$ can be specified in spherical coordinates, expanding the spatial variation along this interface via spherical harmonics. 

Equation \eqref{eq:Lmodecyl} defines a Poisson equation for $\tilde{\phi}_\lambda$, subject to Dirichlet boundary conditions \eqref{eq:dirichlet}. As such, it can be solved by a straightforward application of the finite element method, boundary element method, or directly using the Green's function for Laplace's equation. It is remarkable that \eqref{eq:Lmodecyl} is an inhomogeneous PDE, yet its solutions ultimately obey the homogeneous eigenvalue equation \eqref{eq:eigenb} via \eqref{eq:potendef}. This means we may introduce discontinuities into $\tilde{\bv{E}}_\lambda(\bv{r})$ wherever we please to suit the interface of the target geometry $\theta(\bv{r})$, as opposed to the homogeneous \eqref{eq:eigenb}, where discontinuities typically exist only where embedding geometry $\tilde{\theta}(\bv{r})$ is discontinuous. 

By incorporating knowledge of the location of the discontinuity, we have constructed a more minimal set than \eqref{eq:FBmode},\autocite{chen2020efficient,lobanov2019resonant} reducing the dimensionality of the basis by one. For example, for 2D geometries, \eqref{eq:Lmodecyl} forms a discrete 1D set of modes, in contrast to the Fourier-Bessel basis \eqref{eq:FBmode}, which forms a discrete 2D set of modes. And unlike the Fourier-Bessel basis, we expect the linear combination of \eqref{eq:Lmodecyl} to converge exponentially towards \eqref{eq:divE}, since it incorporates the Dirac-delta component by construction, while the Fourier series expands the smooth variations of the modes on a periodic domain, which is known to converge exponentially.\autocite{strang2007computational} We shall numerically demonstrate this efficiency in Section \ref{sec:numerics}.

One of the few disadvantages of \eqref{eq:Lmodecyl} is that the set $\{\tilde{\phi}_\lambda\}$ is not in general orthogonal with respect to the inner product \eqref{eq:ortho}, for two reasons. Firstly, they all share the same zero eigenvalue of the differential equation \eqref{eq:eigendiff}, so they do not automatically inherit orthogonality from \eqref{eq:ortho}. The definition \eqref{eq:Lmodecyl} also does not furnish an orthogonality relation. Fortunately, this is merely a minor inconvenience, as an orthogonalization procedure such as Gram-Schmidt can be applied if necessary, using \eqref{eq:ortho} as the metric. We choose to use the L\"{o}wden method, since it treats each member symmetrically. This requires evaluating overlap integrals \eqref{eq:ortho} among each mode \eqref{eq:Lmodecyl}, 
\begin{equation}
N_{\nu\mu} \equiv \int \tilde{\bv{E}}_\nu(\bv{r}) \tilde{\theta}(\bv{r}) \tilde{\bv{E}}_\mu(\bv{r})\, d\bv{r},
\end{equation}
and multiplying to obtain orthonormalized longitudinal modes $\tilde{\bv{F}}_\mu$,
\begin{equation}
\tilde{\bv{F}}_\mu = \sum_\nu \tilde{\bv{E}}^\dagger_\nu N^{-\frac{1}{2}}_{\nu\mu}.
\label{eq:Lowden}
\end{equation}

\section{Numerical implementation and examples}
\label{sec:numerics}
The numerical examples of this section serve several purposes, to be a visual illustration of the discontinuous longitudinal basis modes defined by \eqref{eq:Lmodecyl}, to demonstrate their use in the expansion \eqref{eq:expansion} of the target modes \eqref{eq:eigendiff} of an arbitrary geometry, to showcase their convergence properties, and ultimately to solve Maxwell's equations with source terms \eqref{eq:maxwell}. We choose two target 2D geometries for this purpose, a circle and an ellipse.

We first summarize the numerical implementation of the method. This begins by preparing the two sets of basis or embedding modes that emerge from \eqref{eq:eigenb}. A simple embedding geometry is chosen, with analytically known modal fields. For 2D simulations, we choose a circular inclusion of radius $1$, corresponding to $\tilde{\theta}(\bv{r}) = H(1-r)$, where $H(z)$ is the Heaviside step function. We may choose this specific radius without any loss of generality, since only the unitless quantity $kr$ is significant, and $k = \omega/c = 2\pi/\lambda$ can be chosen arbitrarily. 

For circular and spherical embedding geometries, the discontinuous longitudinal basis modes defined by \eqref{eq:Lmodecyl} can be obtained directly using the Green's function for the Laplace operator, supplemented with the method of images to satisfy the Dirichlet boundary condition \eqref{eq:dirichlet}.\autocite{haberman2012applied} The modes are orthonormalized using \eqref{eq:Lowden}. The set of transverse embedding modes and their eigenvalues can be found by solving the well-known dispersion relation for a step-index fiber,
\begin{equation}
\left(\frac{1}{\sqrt{\epsilon_{\tau\tau'}} kB}\frac{J'_\tau(\sqrt{\epsilon_{\tau\tau'}} kB)}{J_\tau(\sqrt{\epsilon_{\tau\tau'}} kB)} - \frac{1}{\sqrt{\epsilon_b} kB}\frac{H'_\tau(\sqrt{\epsilon_b} kB)}{H_\tau(\sqrt{\epsilon_b} kB)}\right) \left(\frac{\sqrt{\epsilon_{\tau\tau'}}}{kB}\frac{J'_\tau(\sqrt{\epsilon_{\tau\tau'}} kB)}{J_\tau(\sqrt{\epsilon_{\tau\tau'}} kB)} - \frac{\sqrt{\epsilon_b}}{kB}\frac{H'_\tau(\sqrt{\epsilon_b} kB)}{H_\tau(\sqrt{\epsilon_b} kB)}\right) = 0.
\label{eq:disprel}
\end{equation}
for which a reliable root search algorithm is available.\autocite{chen2017robust} Due to its continuous rotational symmetry, the modes of \eqref{eq:disprel} can be identified by two so-called quantum numbers: $\tau$ the azimuthal order, and $\tau'$ the radial order, which count the number of nodes along the respective directions. Eigenvalues associated with different azimuthal orders $\tau$ originate from different orders of \eqref{eq:disprel}, while different radial orders $\tau'$ are different roots of the same azimuthal order.

Then, the overlap integrals among the embedding modes \eqref{eq:Vdef} are evaluated for the particular $\theta(\bv{r})$ that defines the target geometry. This forms the matrix eigenvalue equation \eqref{eq:perteig}. Its numerical solution yields the eigenvalues of the target modes and their fields \eqref{eq:eigendiff}. Finally, to solve Maxwell's equations \eqref{eq:maxwell} using GENOME, \eqref{eq:genome} is applied to calculate the total fields produced by any given source configuration.

\subsection{Circular target geometry}
As an initial demonstration, we use the circular embedding modes to obtain the modes of a smaller enclosed circle of radius $0.5$, specifically $\theta(\bv{r}) = H(0.5-r)$. While this is a seemingly trivial application, it allows us to compare the results against one of the few analytical available results, obtained by computing \eqref{eq:disprel} again, this time with $B=0.5$. The target modes \eqref{eq:eigendiff} require a frequency $k$ to be specified, corresponding to an excitation by a monochromatic source with a yet to be defined location. We choose $k = 1$, in units of inverse length.

The cylindrical symmetry of the embedding and target geometries means that all embedding and target modes can be organized by azimuthal quantum number, none of which interact with each other. This symmetry means that its convergence properties are not representative of the general case, so we will defer the convergence analysis until the next geometry. Meanwhile, embedding modes of different radial orders $\tau'$ from \eqref{eq:disprel} do interact to yield the target modes. Since we simulate 2D structures, corresponding to infinite translational symmetry along the third dimension, all modes are either transverse magnetic (TM) or transverse electric (TE). Solving for the TM modes is relatively uninteresting, as no field discontinuities exist and the divergence $\nabla\cdot\bv{E}$ is everywhere zero. Thus, no longitudinal basis modes of any kind are necessary during expansion, so we treat this case only briefly. 

\label{sec:circle}
\begin{figure}[!t]
\begin{center}
\subfloat[$\real(E_z)$]{\begin{tabular}[b]{c}%
\includegraphics[trim={11px 9px 5px 5px},clip]{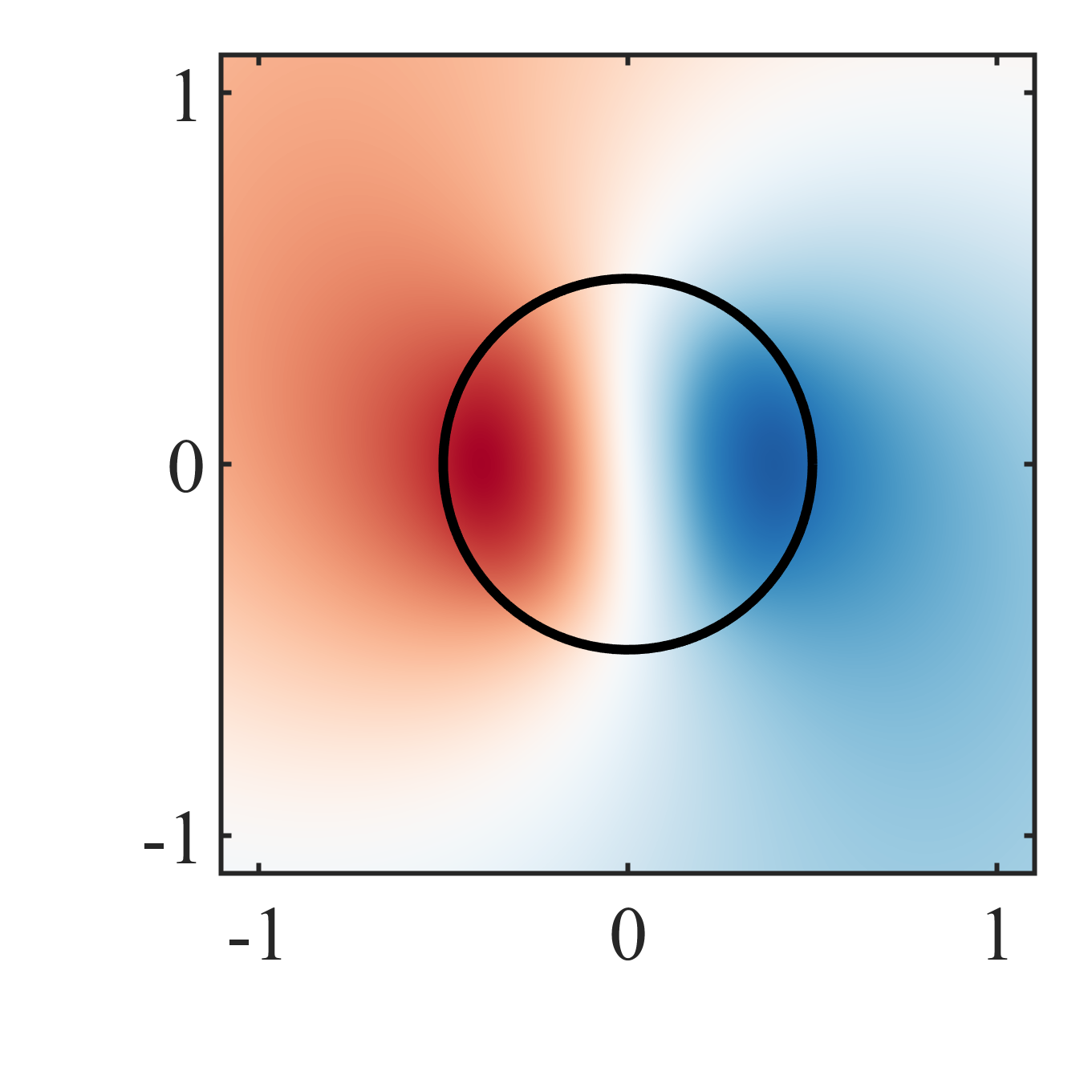}
\end{tabular}}
\subfloat[$\real(E_z)$]{\begin{tabular}[b]{c}%
\includegraphics[trim={11px 9px 5px 5px},clip]{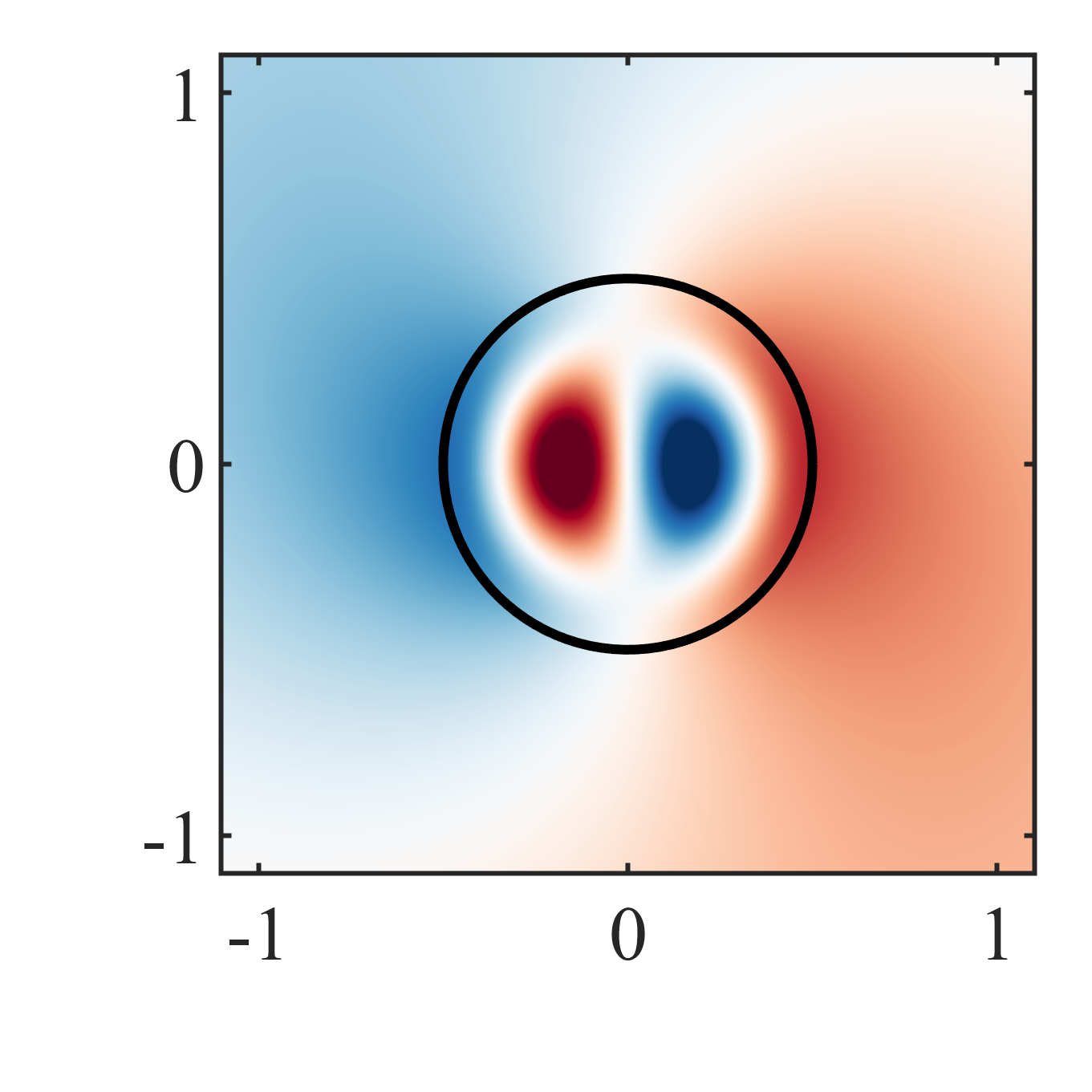}
\end{tabular}}
\subfloat{
\includegraphics[trim={3px 9px 20px 5px},clip]{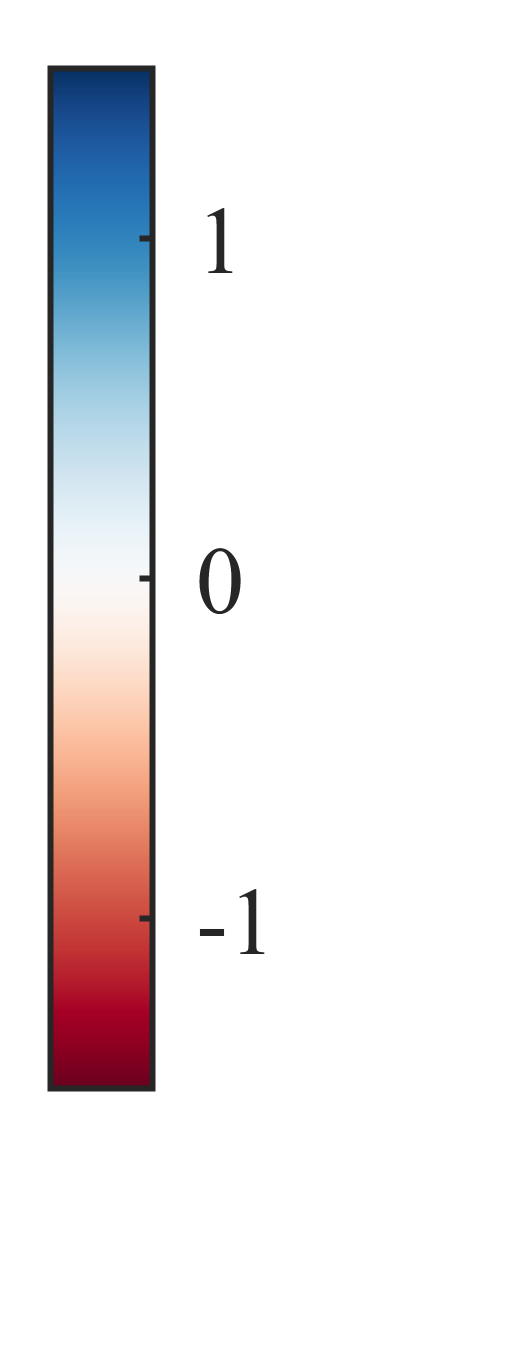}}
\caption{The electric field of the (a) first and (b) second TM modes, with eigenpermittivities are $\epsilon_m = 21.6141 - 2.44876i$ and $\epsilon_m = 120.3096 - 2.3193i$. The real parts are shown. The imaginary parts are not shown, but are identical to the respective real parts after a $90^\circ$ rotation. The black circle corresponds to the boundary of the target cylinder.}
\label{fig:TMmodecirc}
\end{center}
\end{figure}

We show only one example of finding the TM modes by the re-expansion method. Figure \ref{fig:TMmodecirc} shows the two fundamental TM modes of the smaller target cylinder for azimuthal order $\tau=1$, simulated using 50 TM transverse basis modes. The eigenvalues computed by the re-expansion method are $\epsilon_m = 21.61379 - 2.44872i$ and $120.3096 - 2.3193i$, which are accurate to more than $5$ digits compared to the reference values of $\epsilon_m = 21.61374492431008 - 2.44871448053306i$ and $120.3080844540516e - 2.319301692175698i$ , computed directly from the dispersion relation \eqref{eq:disprel}.\autocite{chen2017robust}

\begin{figure}[!t]
\begin{center}
\subfloat[$\real(\tilde{\phi})$]{\begin{tabular}[b]{c}%
\includegraphics[trim={11px 9px 5px 2px},clip]{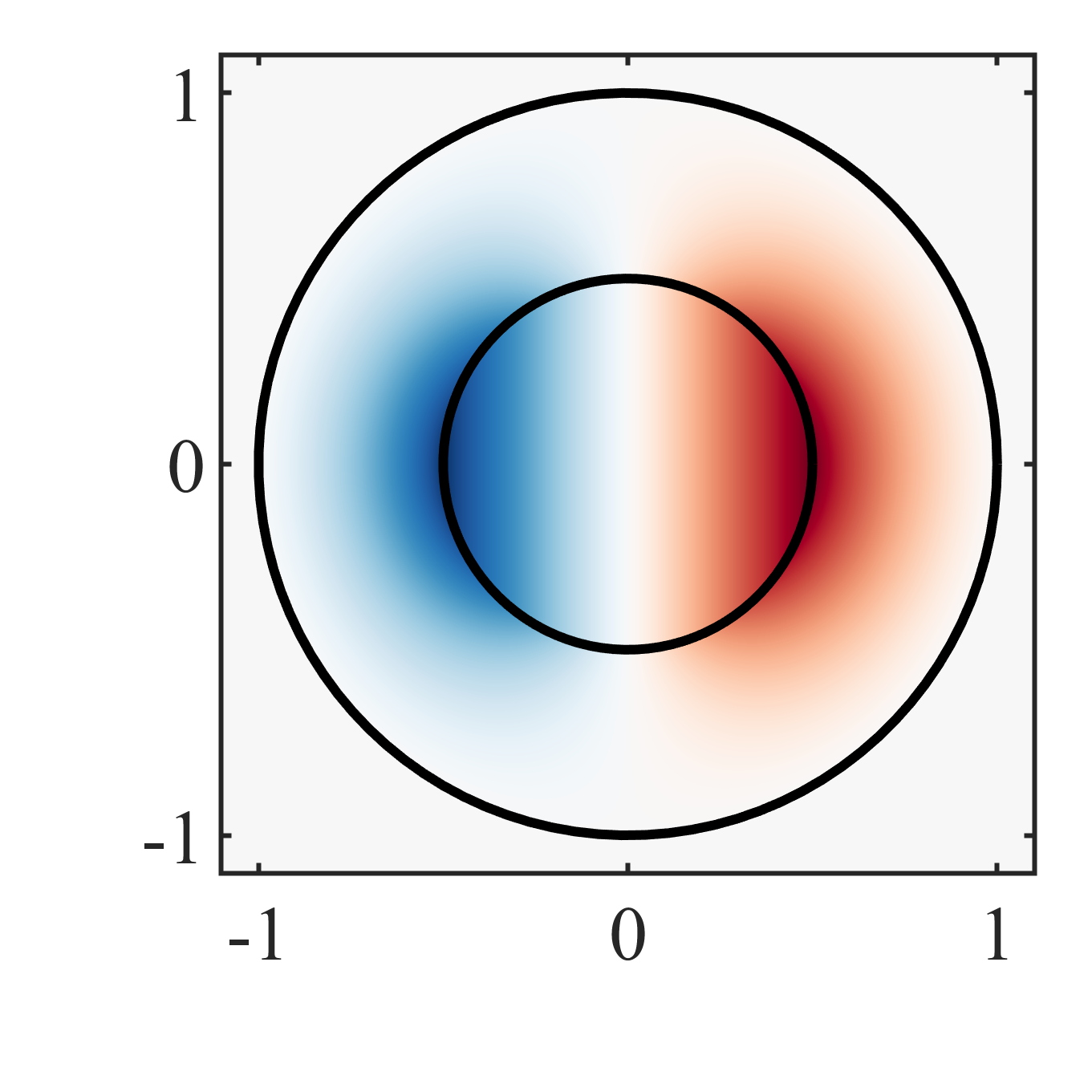}
\end{tabular}}
\subfloat{
\includegraphics[trim={3px 9px 8px 2px},clip]{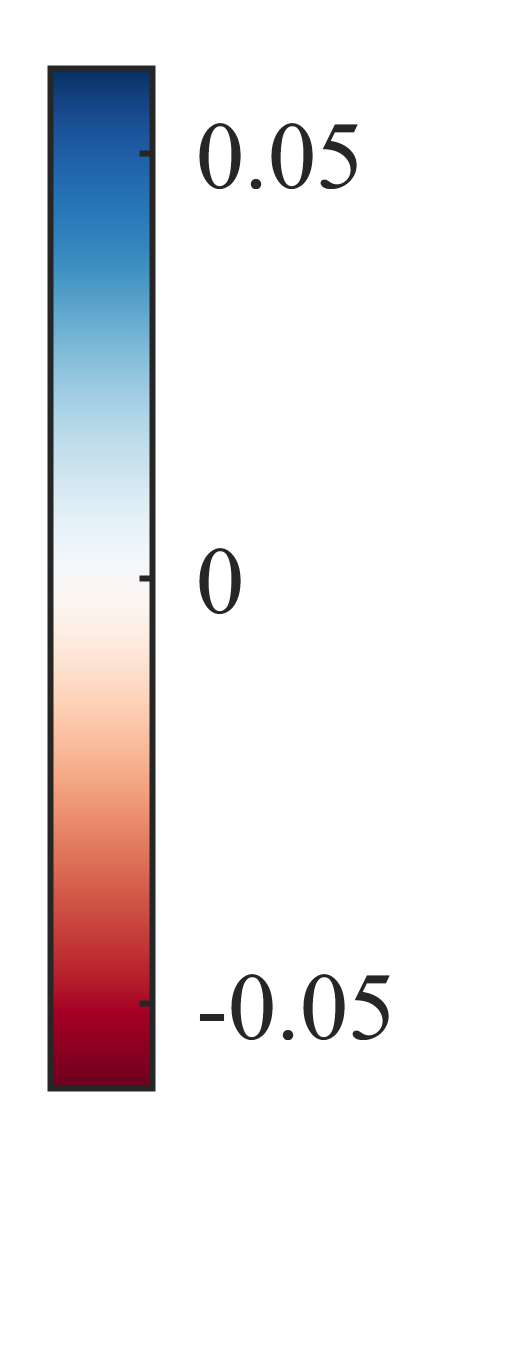}}
\addtocounter{subfigure}{-1}
\subfloat[$\real(\tilde{E}_x)$]{\begin{tabular}[b]{c}%
\includegraphics[trim={11px 9px 5px 2px},clip]{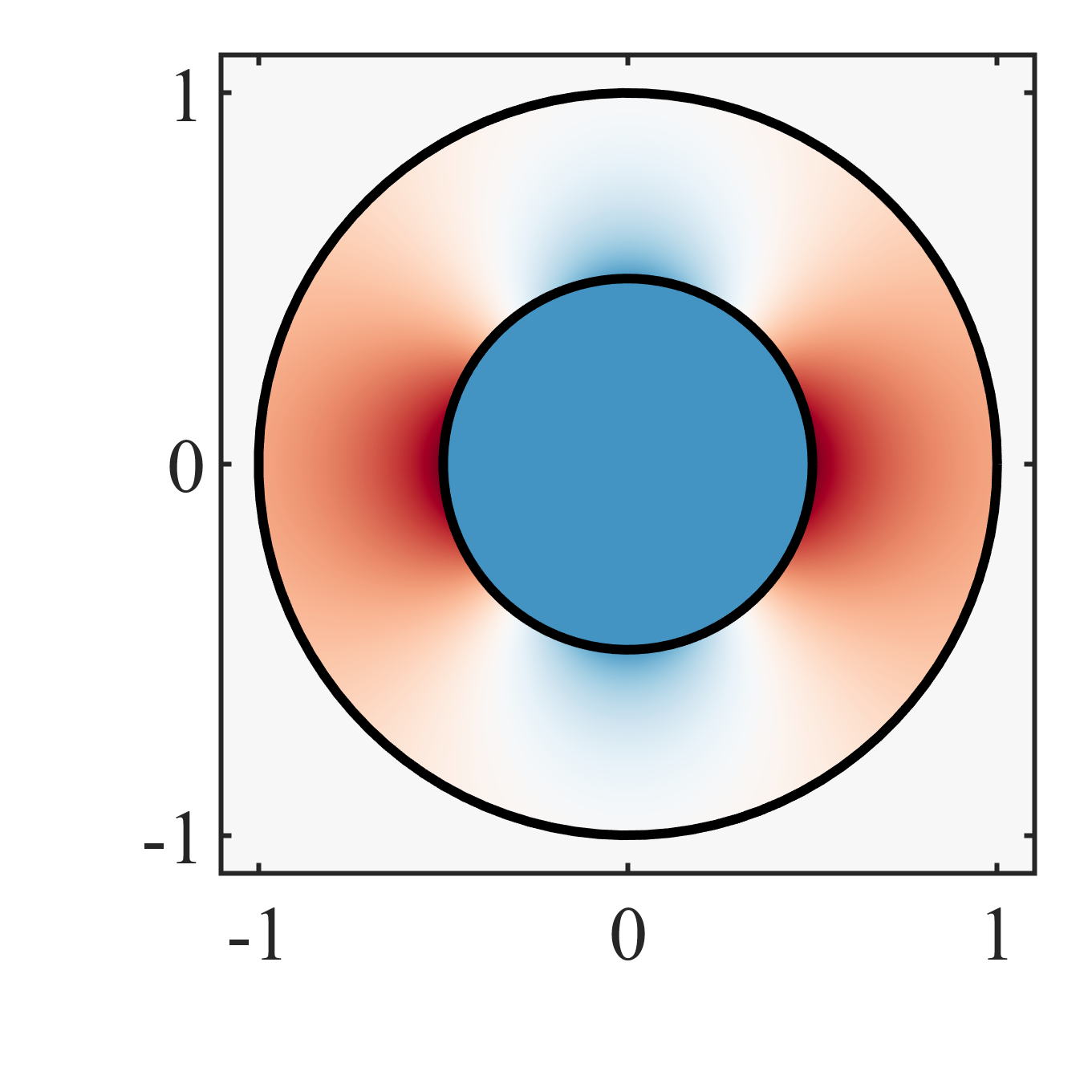}
\end{tabular}}
\subfloat[$\real(\tilde{E}_y)$]{\begin{tabular}[b]{c}%
\includegraphics[trim={11px 9px 5px 2px},clip]{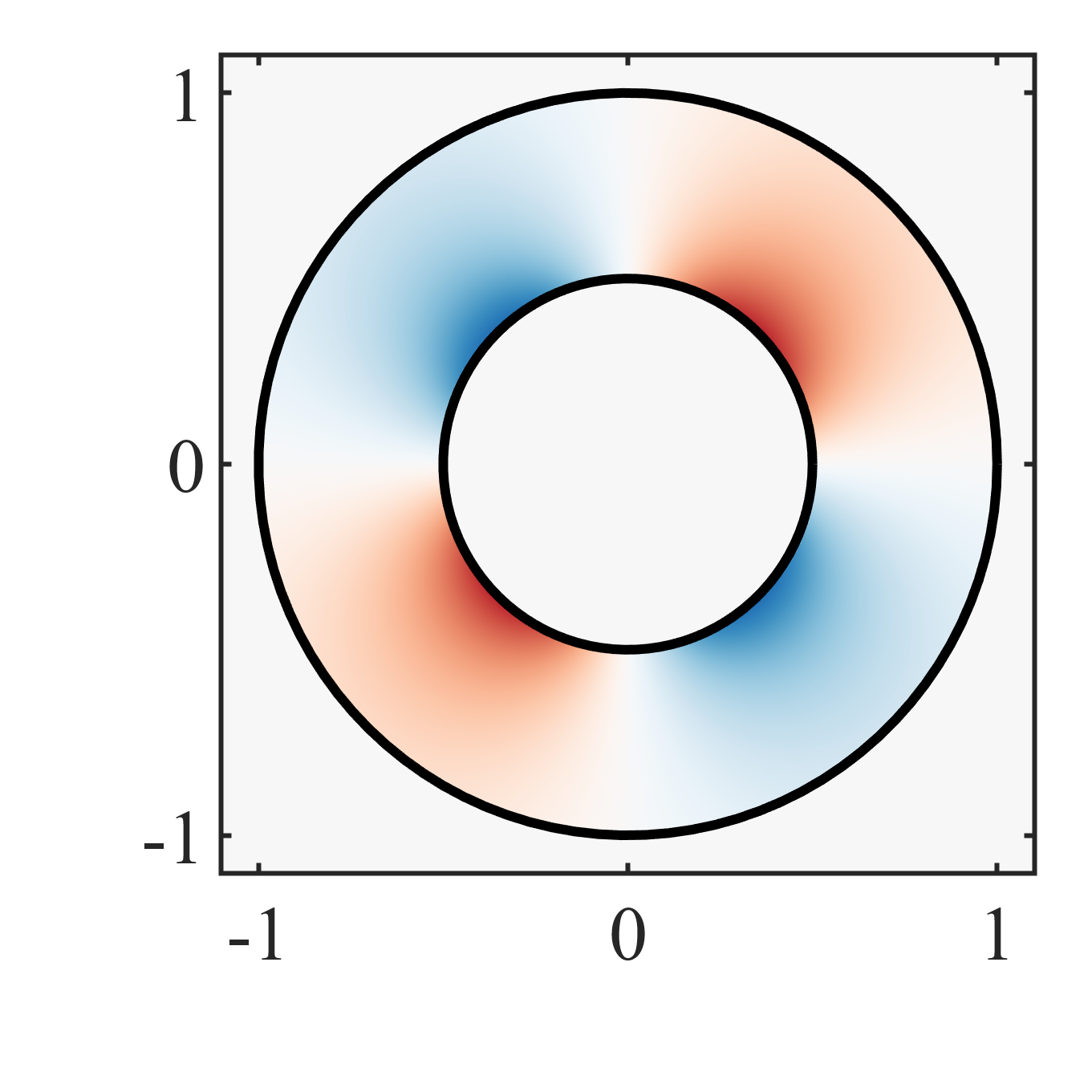}
\end{tabular}}
\subfloat{
\includegraphics[trim={3px 9px 12px 2px},clip]{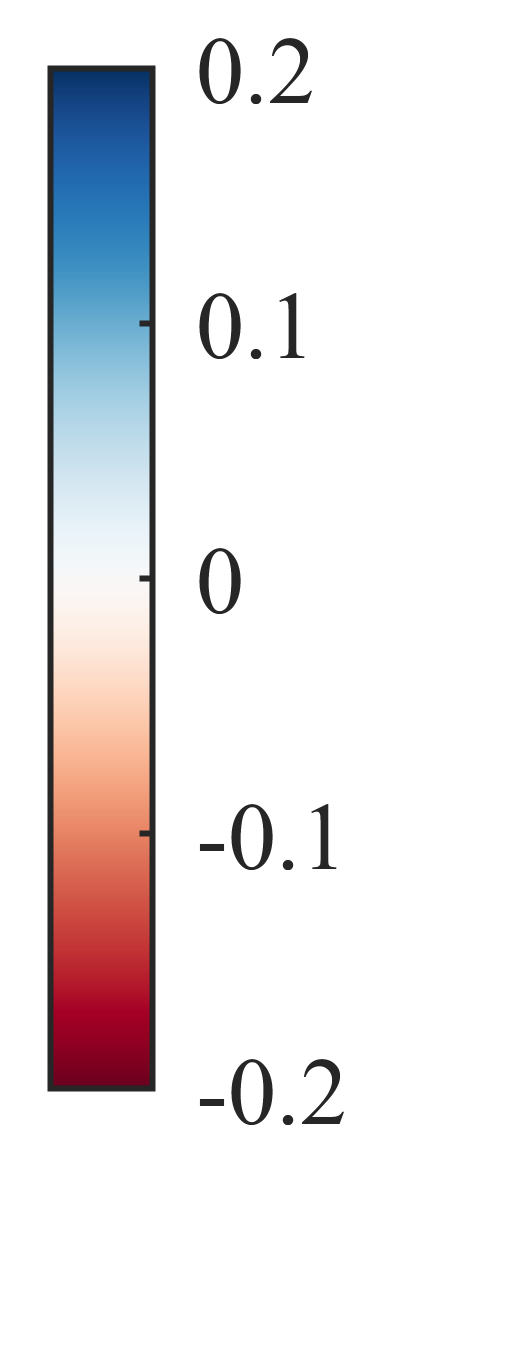}}
\caption{A discontinuous longitudinal mode defined by \eqref{eq:Lmodecyl} for $\lambda=1$. Shown are the real parts of the potential $\tilde{\phi}$ and field components $\tilde{E}_x$ and $\tilde{E}_y$. The imaginary part of $\tilde{\phi}$ is a $90^\circ$ rotation of the real part, and thus the imaginary parts of $\tilde{E}_x$ and $\tilde{E}_y$ are also related to rotations of the real parts. The outer and inner black circles correspond to the basis and target boundaries, respectively.}
\label{fig:longmodecirc}
\end{center}
\end{figure}

Before obtaining the TE target modes, we first discuss some properties of the discontinuous longitudinal modes \eqref{eq:Lmodecyl}. Figure \ref{fig:longmodecirc} shows an example longitudinal mode for $\lambda=1$. In this special case, the integers $\lambda$ correspond to azimuthal orders, since both the basis and target geometries have continuous rotational symmetry. We display the real parts of the potential and electric field components. The potential $\tilde{\phi}$ is zero along the embedding boundary $\partial B$, satisfying the Dirichlet boundary condition \eqref{eq:dirichlet}. It is everywhere continuous. It is derivative-discontinuous normal to the target interface $\partial A$, but is smoothly everywhere else. The modal field $\tilde{\bv{E}} = \nabla\tilde{\phi}$ has corresponding features: its tangential components are zero along the embedding boundary, but there are no restrictions on its normal component here. Fields are discontinuous across the interface $\partial A$, sometimes changing sign. This is important for representing plasmonic modes for example, as these also change sign. Fields of the discontinuous longitudinal basis modes are smoothly varying everywhere else. The fields either decay away from the interface, or remain constant in some special cases. The potential and all fields are zero outside the embedding boundary.

\begin{figure}[!t]
\begin{center}
\subfloat[$\real(E_x)$]{\begin{tabular}[b]{c}%
\includegraphics[trim={11px 9px 5px 5px},clip]{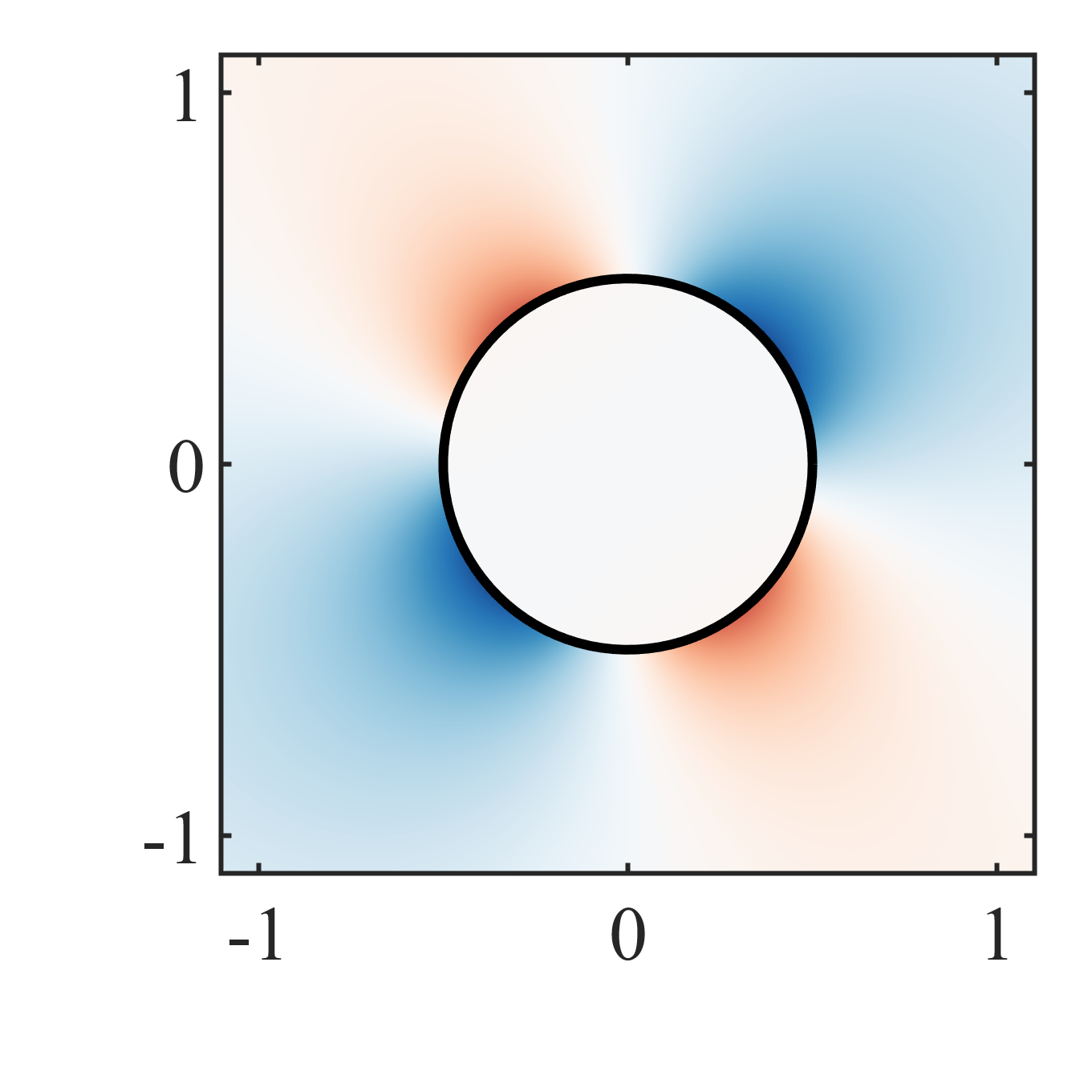}
\end{tabular}}
\subfloat[$\real(E_y)$]{\begin{tabular}[b]{c}%
\includegraphics[trim={11px 9px 5px 5px},clip]{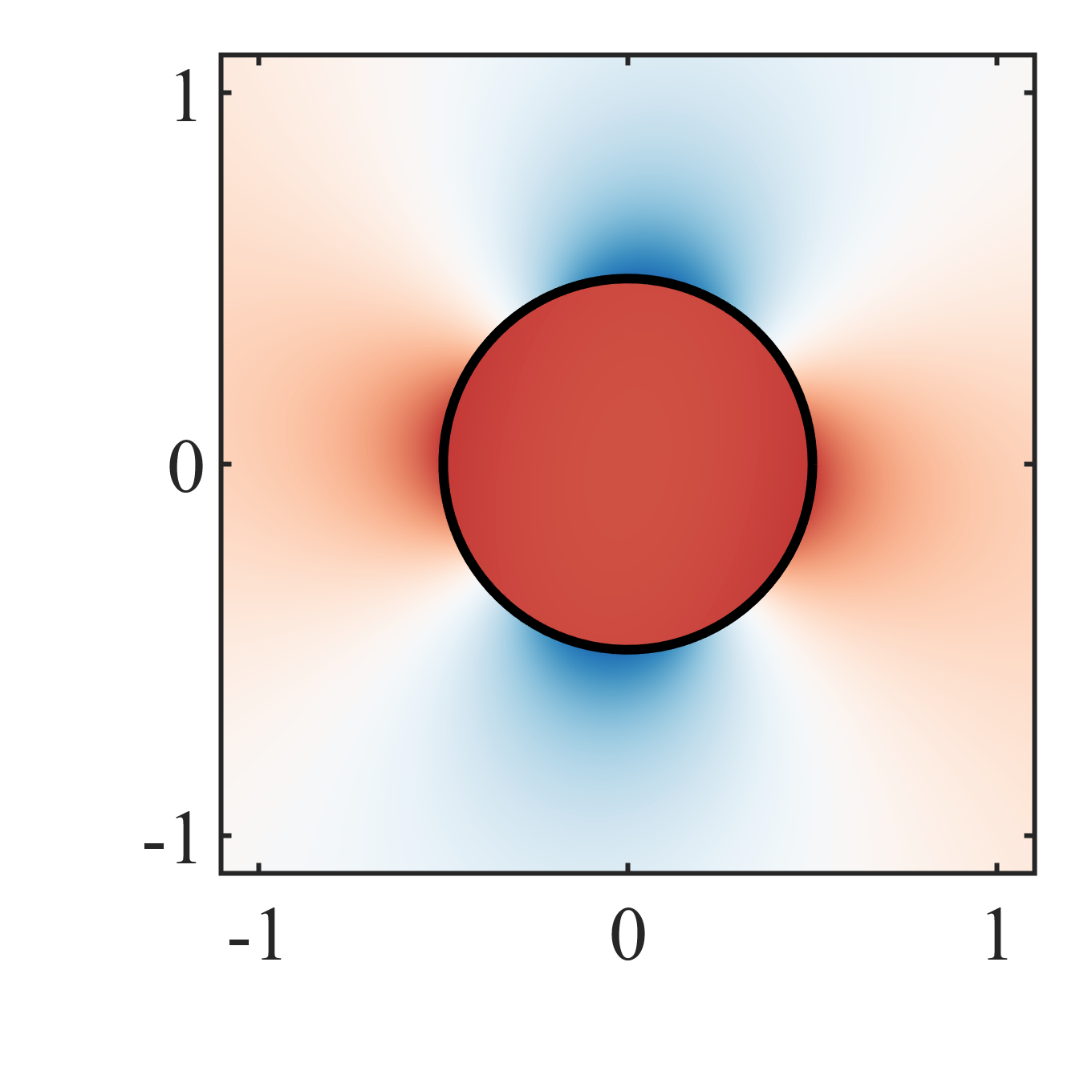}
\end{tabular}}
\subfloat{
\includegraphics[trim={3px 9px 12px 5px},clip]{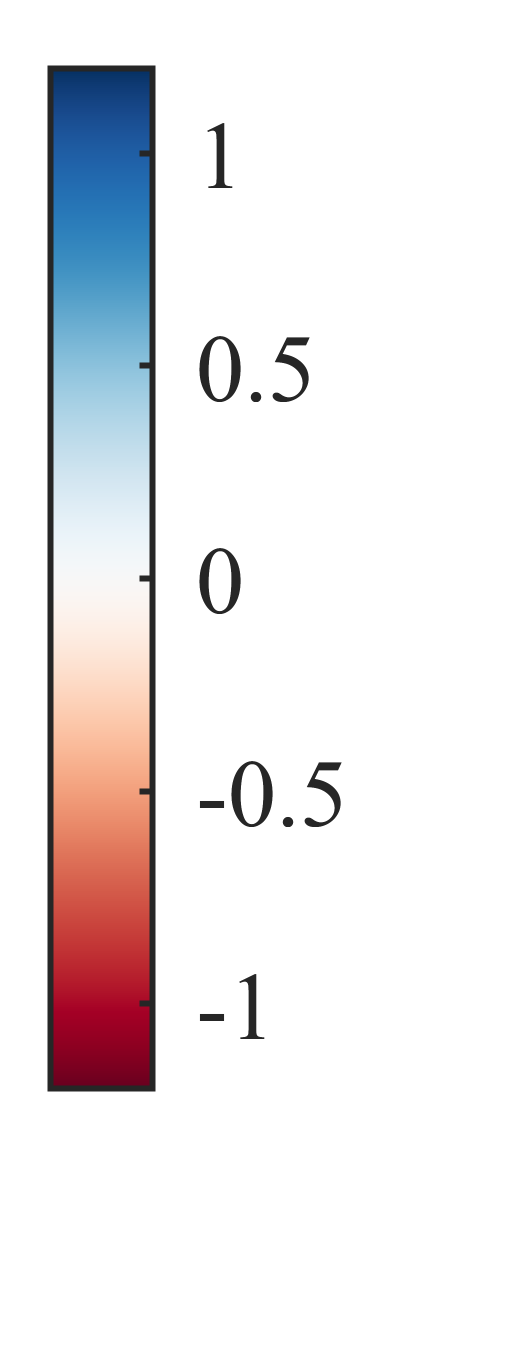}}\\
\addtocounter{subfigure}{-1}
\subfloat[$\real(E_x)$]{\begin{tabular}[b]{c}%
\includegraphics[trim={11px 9px 5px 5px},clip]{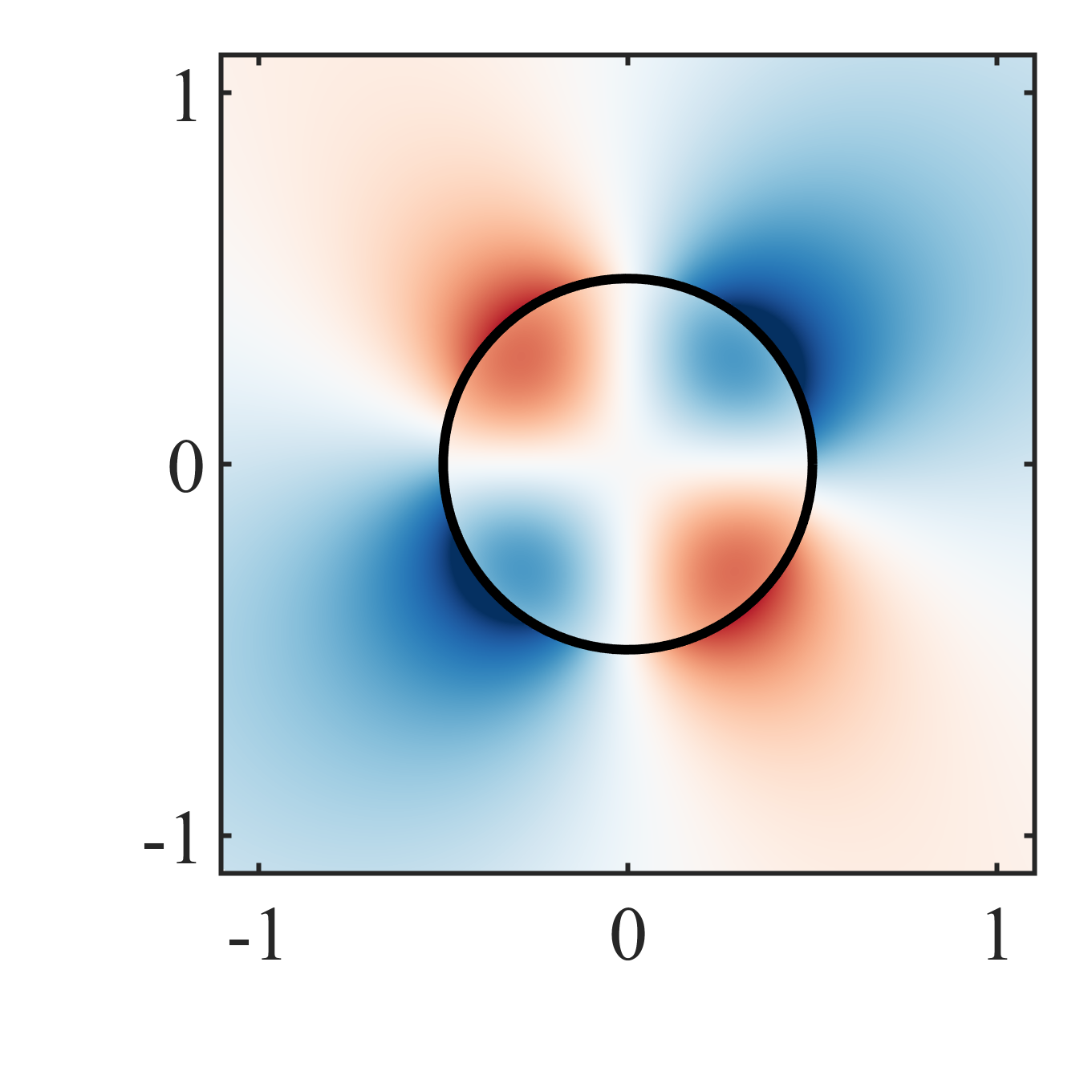}
\end{tabular}}
\subfloat[$\real(E_y)$]{\begin{tabular}[b]{c}%
\includegraphics[trim={11px 9px 5px 5px},clip]{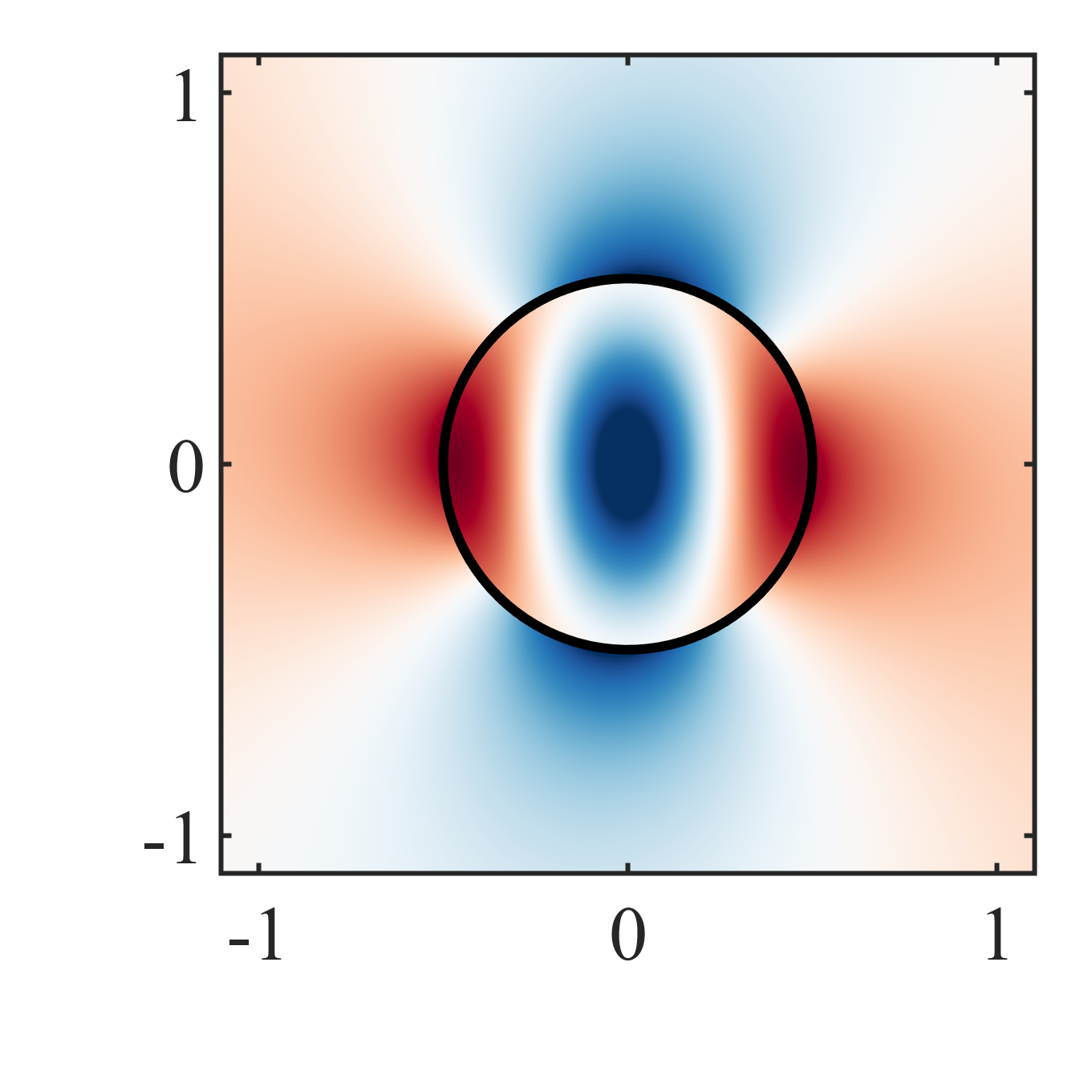}
\end{tabular}}
\subfloat{
\includegraphics[trim={3px 9px 12px 5px},clip]{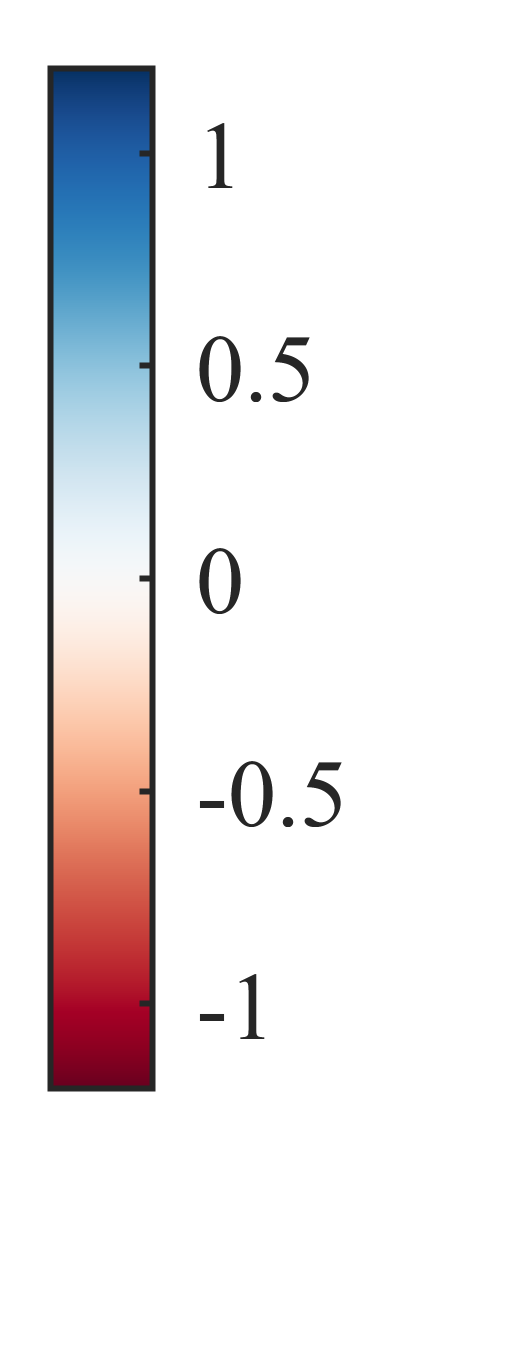}}
\caption{(a), (b) As in Figure \ref{fig:TMmodecirc}, but showing instead the brightest plasmonic mode of the circle, displaying the real parts of $E_x$ and $E_y$, with eigenpermittivity $\epsilon_m = -1.175664 - 0.454290i$. (c), (d) Shows the fundamental dielectric TE mode, with $\epsilon_m = 56.4856 - 0.8179i$.}
\label{fig:plasmodecirc}
\end{center}
\end{figure}

We now present numerical examples of using a discontinuous longitudinal embedding mode to generate TE modes of the smaller enclosed circle. Displayed in Figure \ref{fig:plasmodecirc} is the plasmonic mode and fundamental dielectric mode, both of azimuthal order $\tau = 1$. A major component of both these modes is the same $\lambda=1$ discontinuous longitudinal mode of Figure \ref{fig:longmodecirc}. Indeed, failure to include any longitudinal modes will never result in a mode with the correct eigenvalue regardless of the number of transverse embedding modes used. The simulated eigenpermittivities are $\epsilon_m = -1.1756666 - 0.4542910i$ and $56.4804 - 0.8178i$, which compares with the reference values $\epsilon_m = -1.175666945325108 - 0.454291223574987i$ and $56.480144191790039 - 0.817845963134636i$. Shown are the real and imaginary parts of the $E_x$ and $E_y$ fields. The fields of the plasmonic mode changes sign at the interface, as expected, which the $\lambda = 1$ discontinuous longitudinal mode plays a considerable role in accurately reproducing. It is also important for reproducing the discontinuity at the interface of the dielectric mode, even though no sign change occurs.

Since these are TE modes, the $E_z$ field should be identically zero, which the re-expansion method reproduces to numerical precision. The modes show no obvious discontinuities at the artificial embedding boundary $\partial B$, even though the constituent longitudinal and transverse embedding modes are discontinuous here. This is one indication of the accuracy of the modes, since only the correct combination of both longitudinal and transverse embedding modes produces a smooth field here, since otherwise a halo would appear at $r=1$. The mode has been normalized.

The discontinuous longitudinal modes are in some sense the ideal basis modes for piecewise uniform geometries, which is exemplified by this special case of a circular target geometry. For each azimuthal order, only one longitudinal mode \eqref{eq:Lmodecyl} exists, while an infinite number of transverse modes exists of different radial orders $\tau'$. Since modes of different azimuthal orders do not intermix, only a single discontinuous longitudinal embedding mode is necessary to generate each target mode of the smaller enclosed circle, even though an infinite number of transverse embedding modes is in principle necessary. Furthermore, this one discontinuous longitudinal embedding mode is capable of generating all the target modes of a single azimuthal order, regardless of whether these target modes are plasmonic or dielectric in character, and regardless of other parameters such as frequency. This is apparent from the similarity between the $\lambda=1$ discontinuous longitudinal mode of Figure \ref{fig:longmodecirc} and the two modes of Figure \ref{fig:plasmodecirc}, in particular by comparing the field patterns in the annular region between the inner and outer circles of Figure \ref{fig:longmodecirc}. These remarkable facts are due to the dimension reduction inherent to the discontinuous longitudinal modes.

\subsection{Elliptical target geometry}
\label{sec:ellipse}
\begin{figure}[!t]
\begin{center}
\subfloat[$\real(\tilde{\phi})$]{\begin{tabular}[b]{c}%
\includegraphics[trim={11px 9px 5px 2px},clip]{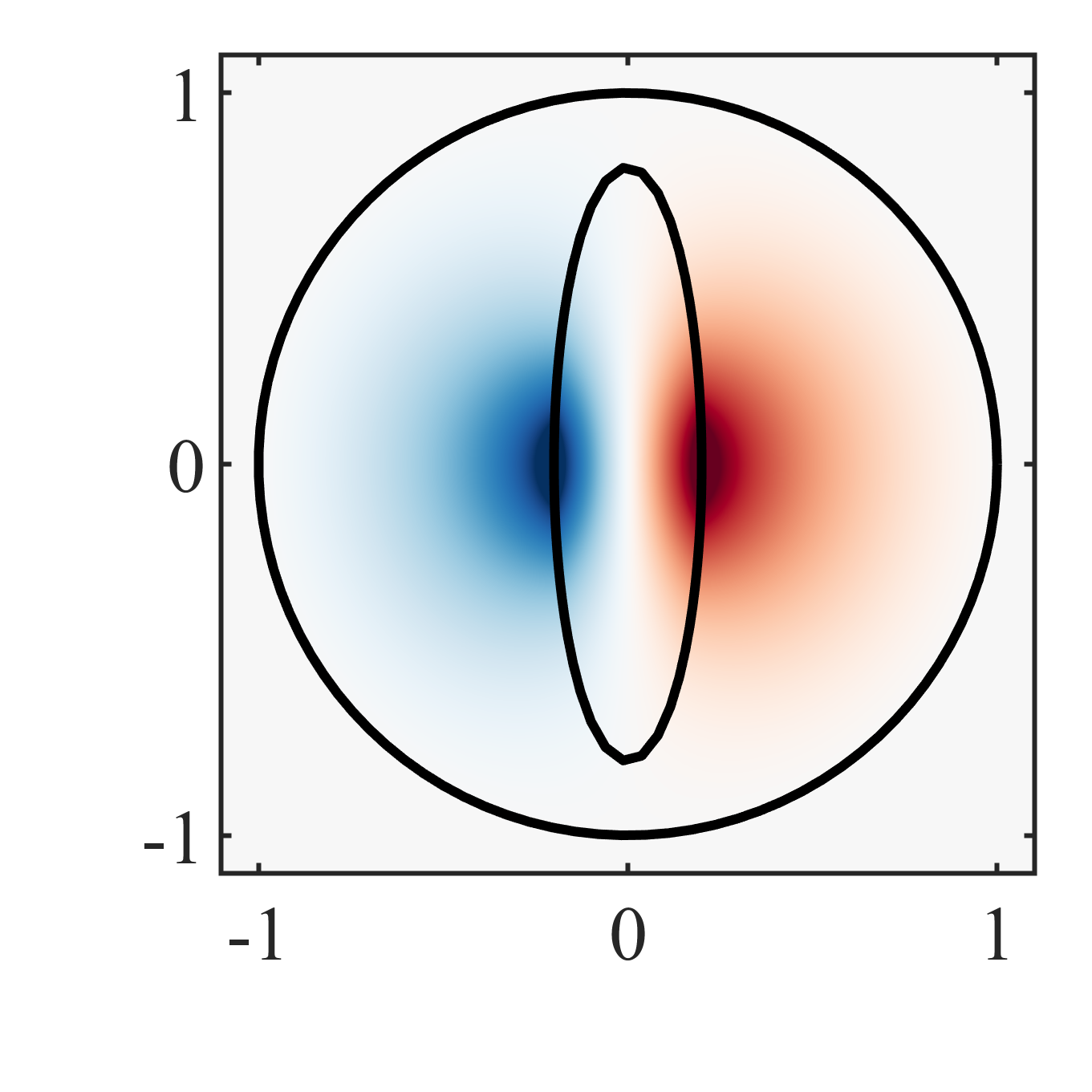}
\end{tabular}}
\subfloat{
\includegraphics[trim={3px 9px 8px 2px},clip]{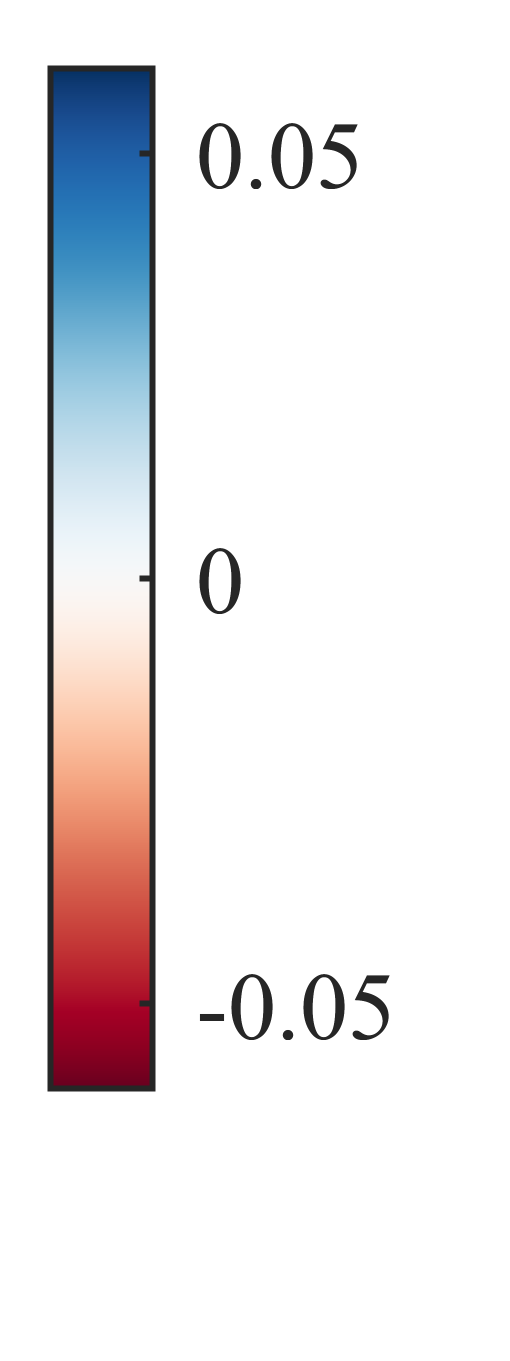}}
\addtocounter{subfigure}{-1}
\subfloat[$\real(\tilde{E}_x)$]{\begin{tabular}[b]{c}%
\includegraphics[trim={11px 9px 5px 2px},clip]{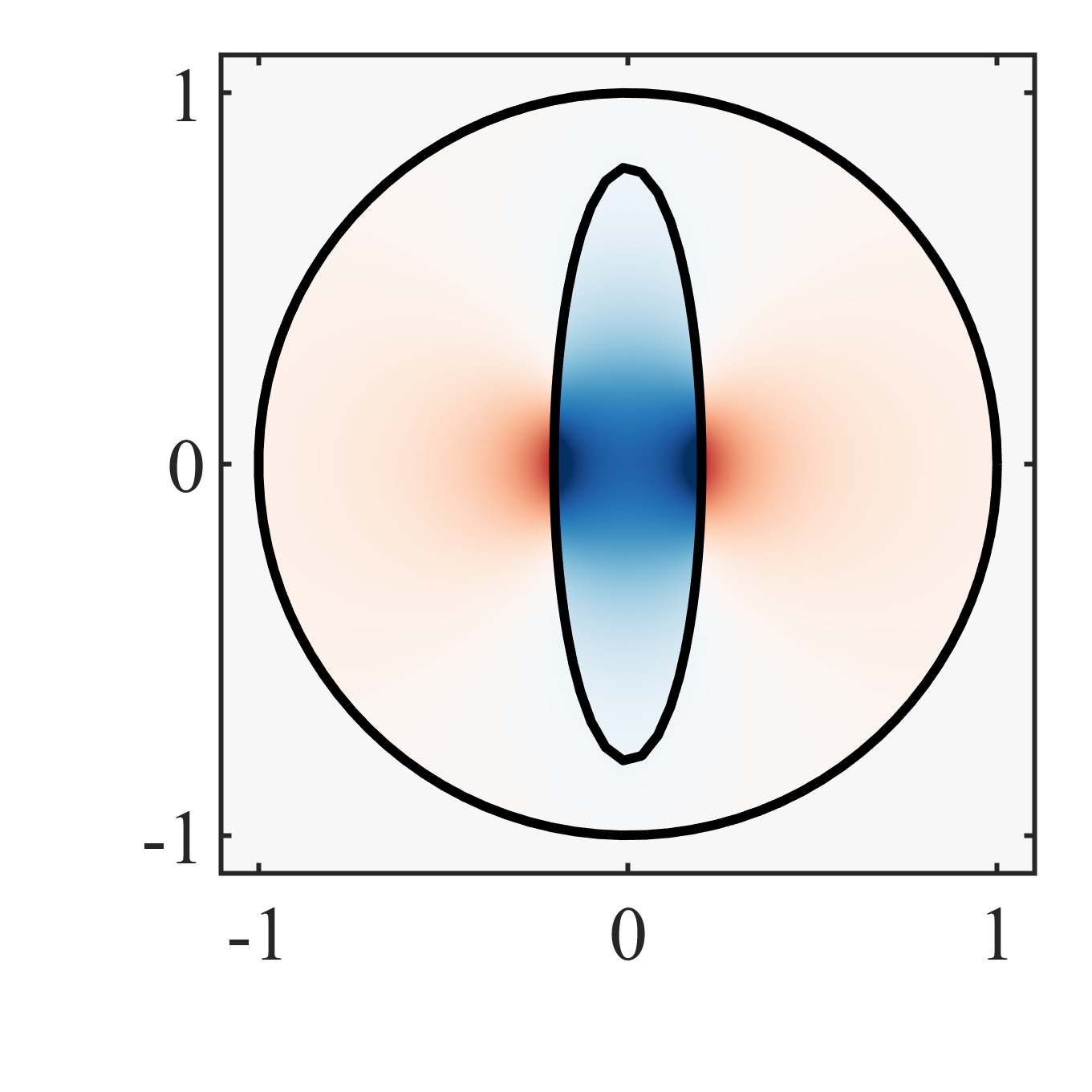}
\end{tabular}}
\subfloat[$\real(\tilde{E}_y)$]{\begin{tabular}[b]{c}%
\includegraphics[trim={11px 9px 5px 2px},clip]{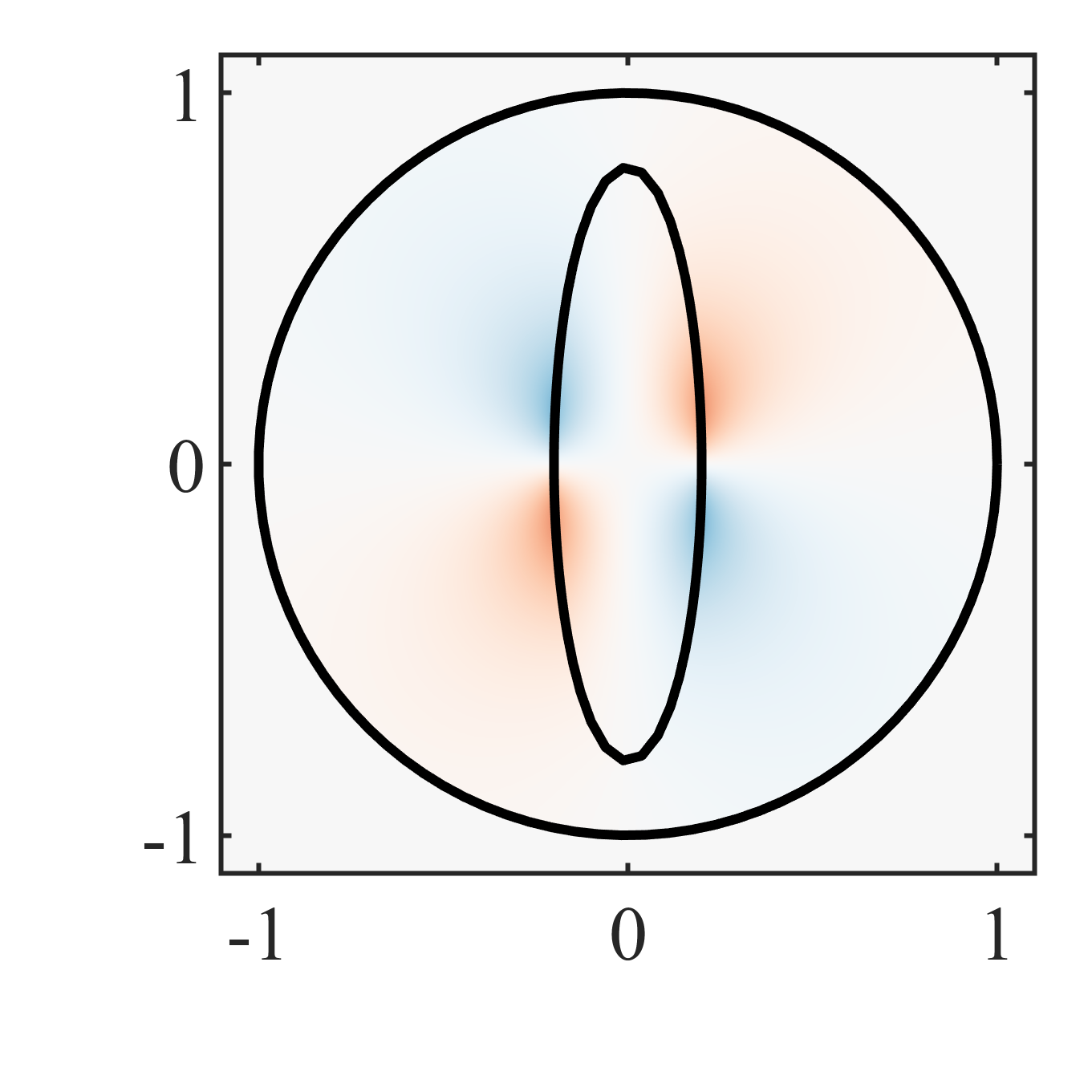}
\end{tabular}}
\subfloat{
\includegraphics[trim={3px 9px 12px 2px},clip]{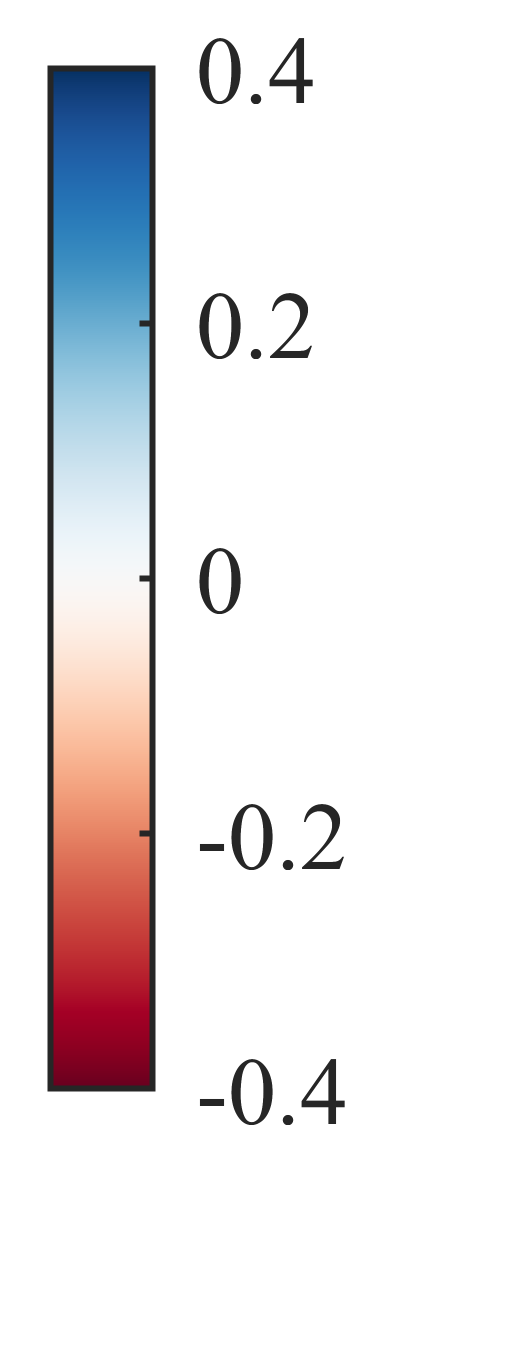}}
\caption{As in Figure \ref{fig:longmodecirc}, showing a discontinuous longitudinal mode defined by \eqref{eq:Lmodecyl} for $\lambda=1$. Shown are the real parts of the potential $\tilde{\phi}$ and field components $\tilde{E}_x$ and $\tilde{E}_y$. The outer circle indicates the basis boundary, while the inner ellipse is the target geometry.}
\label{fig:longmodeelli}
\end{center}
\end{figure}

We now apply the re-expansion method with discontinuous longitudinal modes to find the modes of an elliptical geometry pictured in Figure \ref{fig:embedding}, defined by $\theta(\bv{r}) = H(1-(x/a)^2-(y/b)^2)$. The semi-major axis is chosen to be $a=0.4$ and the semi-minor axis is $b=0.1$. It is excited by a monochromatic source, oscillating at $k=1$, in units of inverse length. This is a subwavelength structure, where the ratio of the length of the ellipse along the major axis to the wavelength is approximately $0.125$.

To expand the target modes, we employ the same transverse embedding modes, found via \eqref{eq:disprel}, as in the previous example. But the discontinuity of the longitudinal embedding modes \eqref{eq:Lmodecyl} is now specific to the elliptical interface. We plot an example of these modes in Figure \ref{fig:longmodeelli}. Again, the potential $\tilde{\phi}_\lambda$ and electric field components are shown, for order $\lambda = 1$. Their behavior resembles the mode of Figure \ref{fig:longmodecirc}. The various boundary conditions are satisfied. Correspondingly, fields are discontinuous across the elliptical interface, and are otherwise smooth.

\begin{figure}[!t]
\begin{center}
\subfloat[$\real(E_x)$]{\begin{tabular}[b]{c}%
\includegraphics[trim={11px 9px 5px 2px},clip]{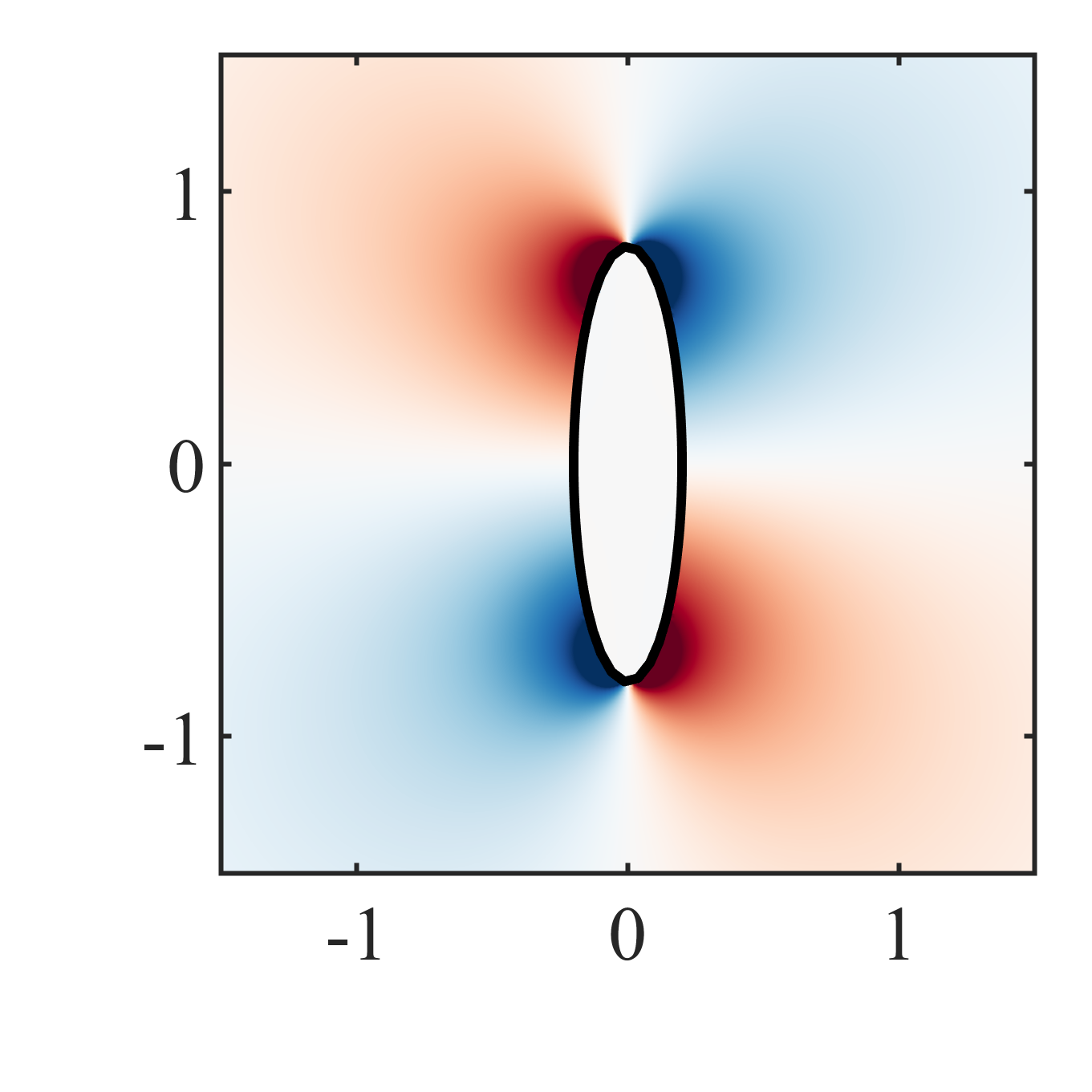}
\end{tabular}}
\subfloat[$\real(E_y)$]{\begin{tabular}[b]{c}%
\includegraphics[trim={11px 9px 5px 2px},clip]{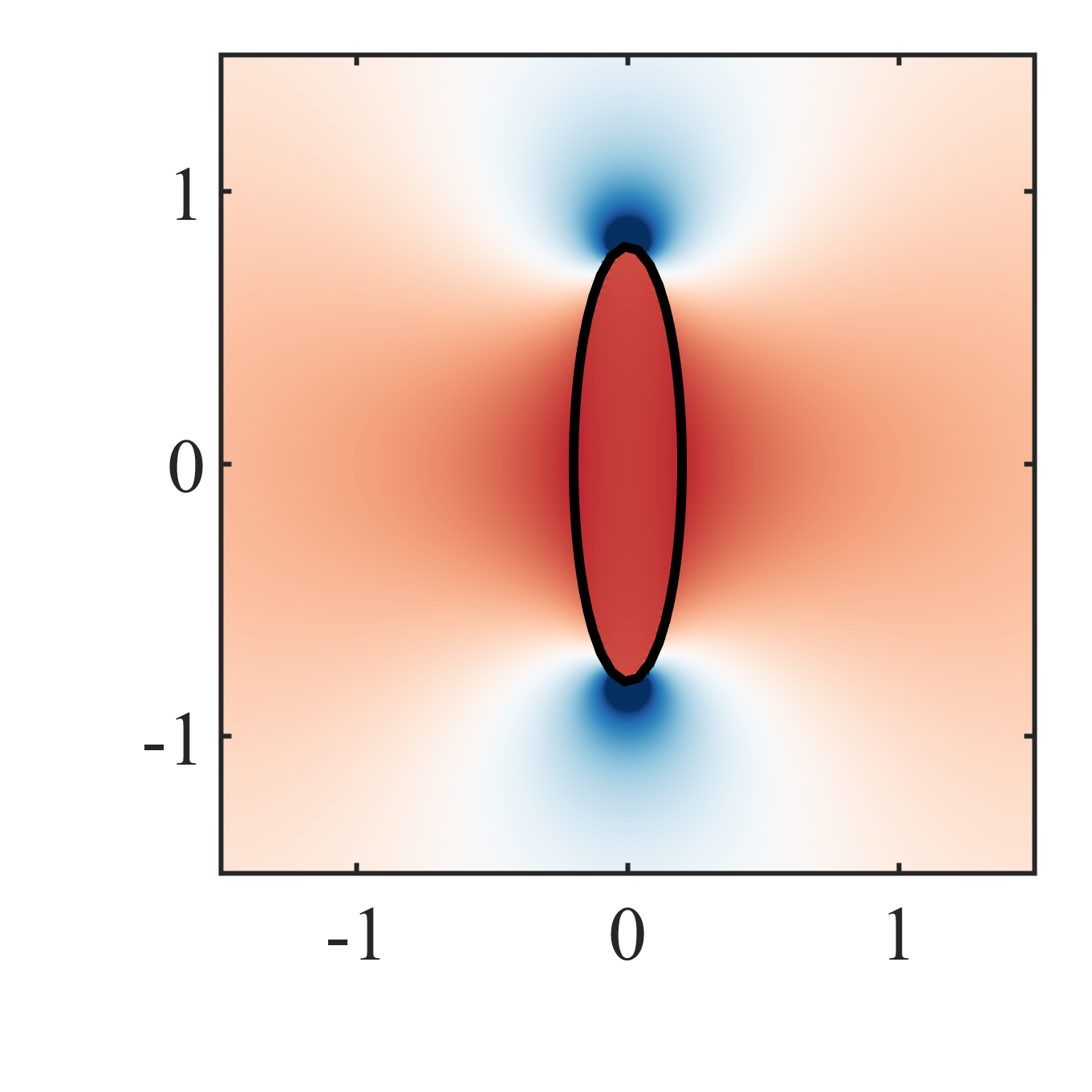}
\end{tabular}}
\subfloat{
\includegraphics[trim={3px 9px 15px 2px},clip]{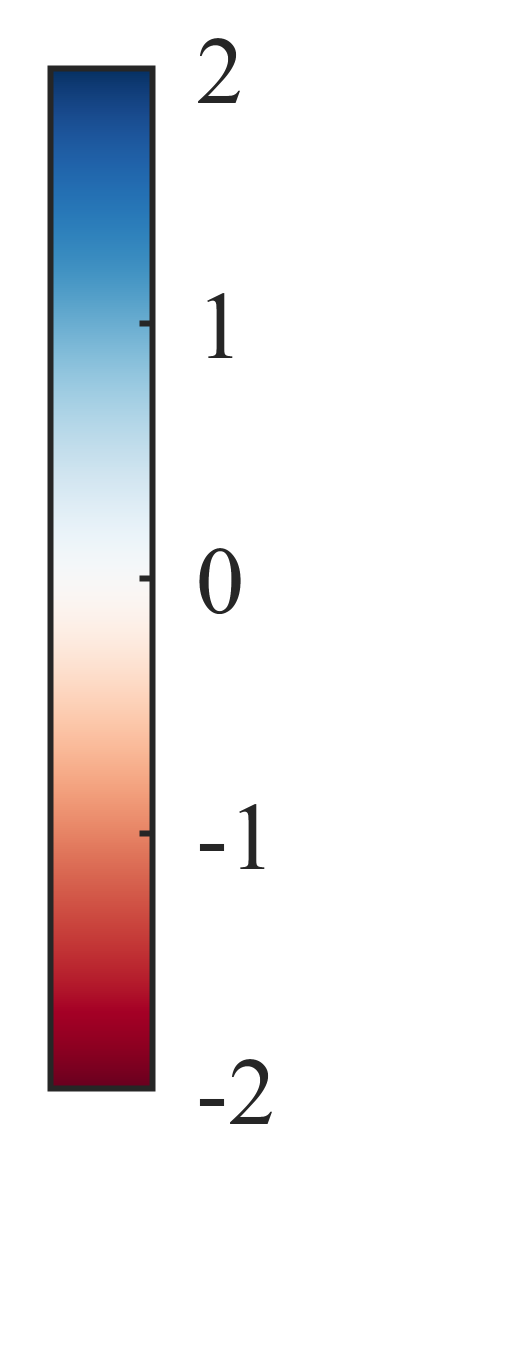}}
\caption{The brightest plasmonic mode of the ellipse, showing the real parts of field components $E_x$ and $E_y$. The eigenpermittivity is $\epsilon_m = -4.78991 - 2.33514i$.}
\label{fig:plasmodeelli}
\end{center}
\end{figure}

\begin{figure}[!t]
\begin{center}
\includegraphics{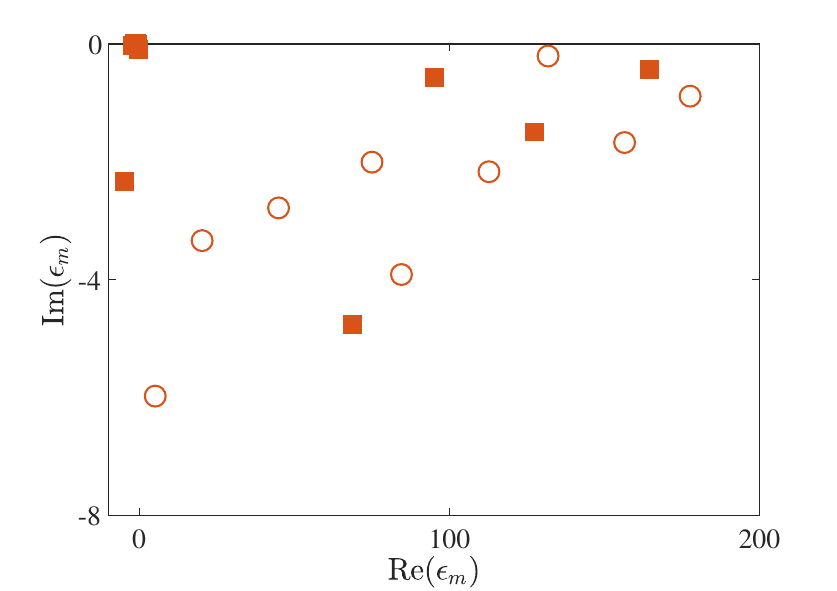}
\caption{Eigenpermittivites of the modes of the ellipse, plotting the real and imaginary parts. TM modes are indicated by hollow round markers, and TE modes by solid square markers.}
\label{fig:epselli}
\end{center}
\end{figure}

Now, overlap integrals \eqref{eq:Vdef} between modes of all azimuthal orders are performed, corresponding to all values of $\lambda$ and $\tau$ up to truncation. In other words, the different orders do not decouple, unlike the circular example of Section \ref{sec:circle}. For demonstration purposes, we initially employed many embedding modes, 100 longitudinal and 5000 transverse, to ensure that a highly accurate reference result is obtained. After diagonalizing \eqref{eq:perteig}, we generate in a single simulation many hundreds of usable modes of the ellipse, both plasmonic and dielectric, and both TE and TM. In particular, a large number of accurate plasmonic modes were found, without any missing modes. Their eigenvalues are plotted in Figure \ref{fig:epselli}. The number of accurate higher-order plasmonic modes we find seems to be limited only by the number of longitudinal modes employed.

We now briefly discuss the physical significance of the eigenvalues $\epsilon_m$. A negative $\real(\epsilon_m)$ indicates a plasmonic mode, with modal fields that change sign at the elliptical interface, while a positive $\real(\epsilon_m)$ indicates a dielectric mode with no such sign change. All modes have negative $\imag(\epsilon_m)$, with more negative values indicating brighter modes that radiate more energy into the far-field. In Figure \ref{fig:epselli}, all plasmonic modes are clustered near the origin, except the bright plasmonic mode at $-4.78991 - 2.33514i$, which we shall call the fundamental plasmonic mode. The dielectric modes are distributed across the figure, with few discernible trends. The only trend is that modes with small $\real(\epsilon_m)$ tend to have more negative $\imag(\epsilon_m)$. This is because modes with large $\real(\epsilon_m)$ have more nodes in the angular direction, leading to greater trapping of energy, an effect known as the whispering gallery mode.

We now focus on the fundamental plasmonic mode. Figure \ref{fig:plasmodeelli} plots its fields, which is dipolar with an oscillation along the long axis. Shown are the real and imaginary parts of the $E_x$ and $E_y$ fields. As a plasmonic mode, visible sign changes at the interface of these modes, which is well-represented by the discontinuous longitudinal embedding modes. As with all plasmonic modes, the mode is TE, so the $E_z$ field should be identically zero, which the re-expansion method reproduces to numerical precision. The mode shows no obvious discontinuities at the artificial basis boundary, which comports with the behavior of the plasmonic mode of Figure \ref{fig:plasmodecirc}. In part because of its importance, the convergence tests of Section \ref{sec:convergence} are performed on this mode.

\begin{figure}[!t]
\begin{center}
\subfloat[$\real(E_x)$]{\begin{tabular}[b]{c}%
\includegraphics[trim={11px 9px 5px 2px},clip]{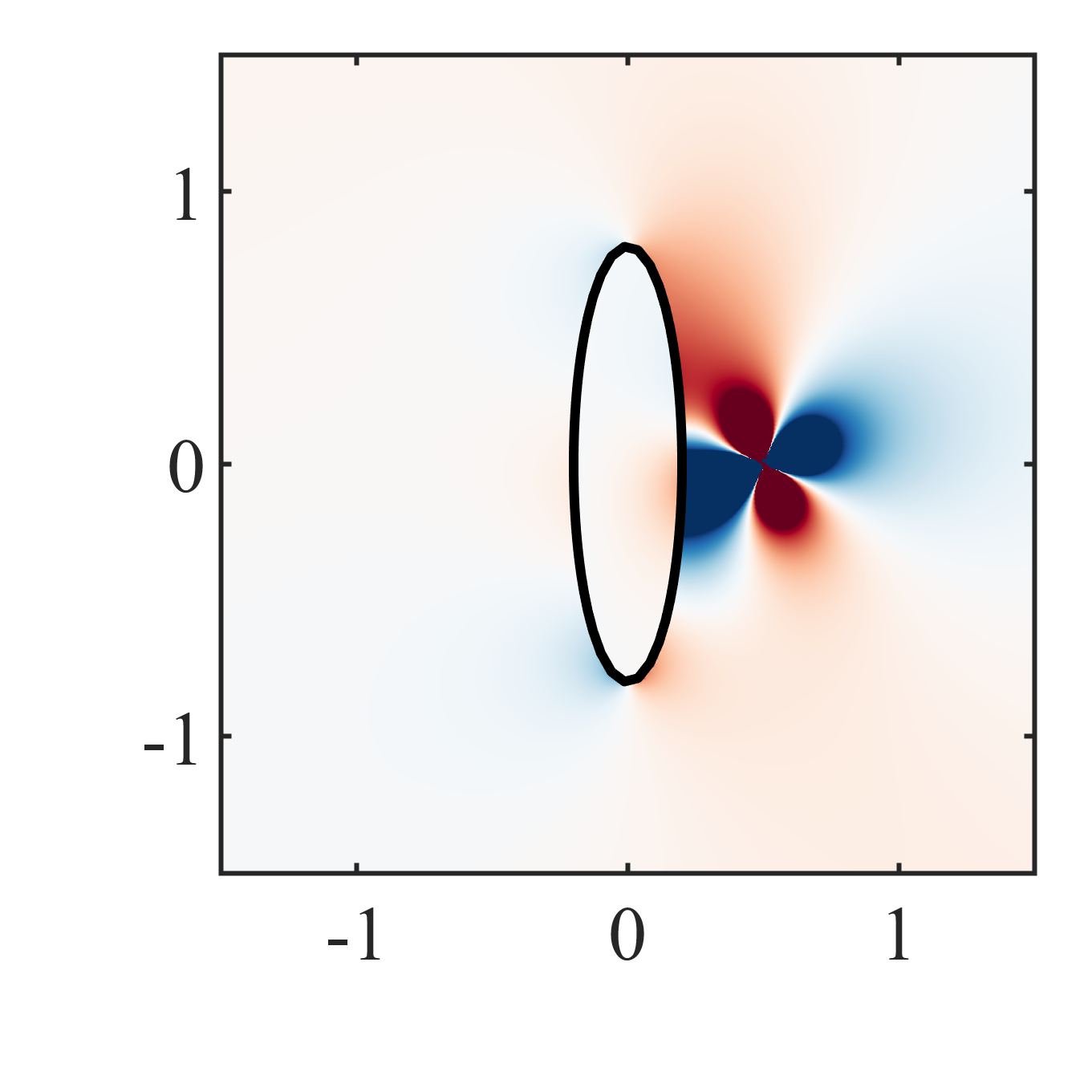}
\end{tabular}}
\subfloat[$\real(E_y)$]{\begin{tabular}[b]{c}%
\includegraphics[trim={11px 9px 5px 2px},clip]{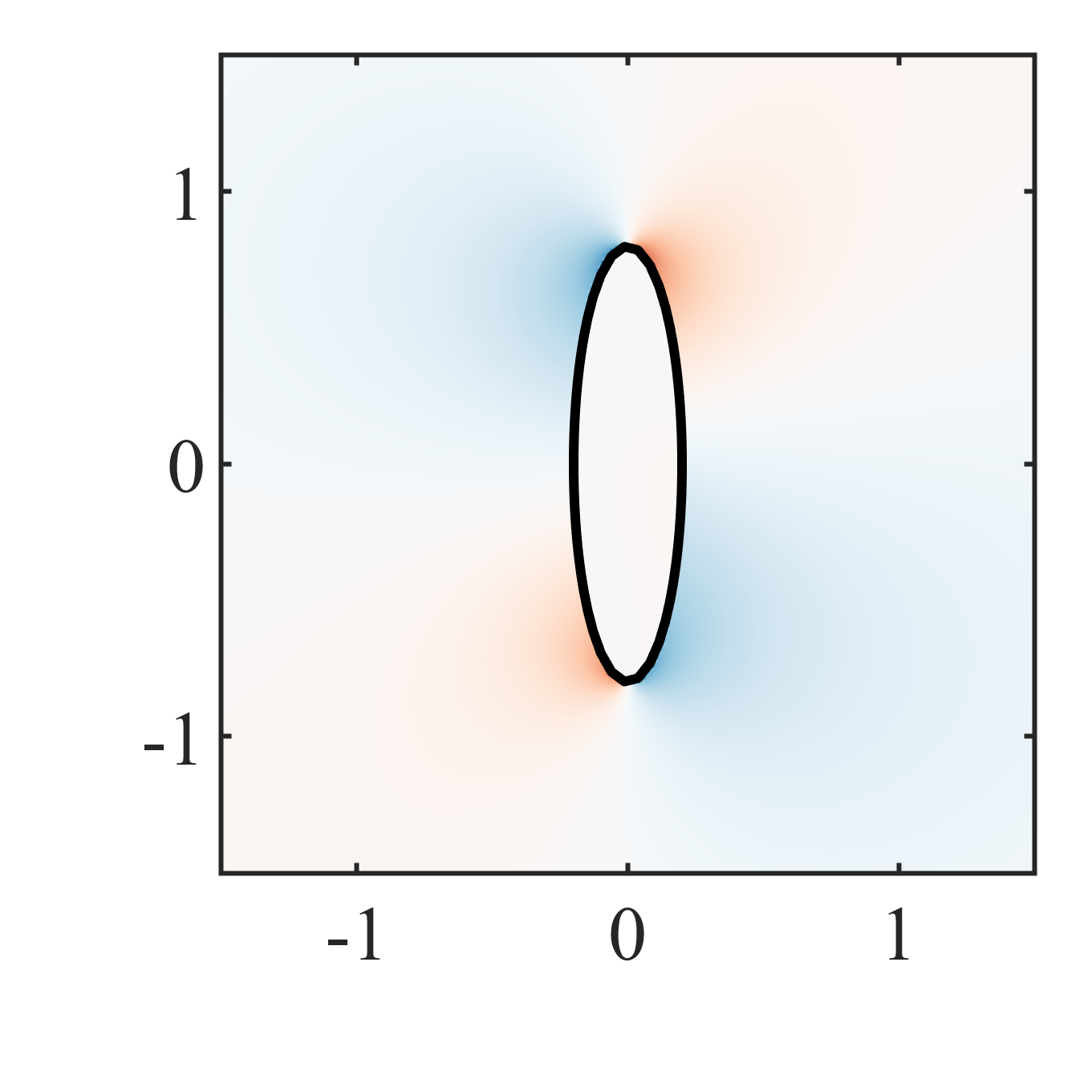}
\end{tabular}}
\subfloat{
\includegraphics[trim={3px 9px 12px 2px},clip]{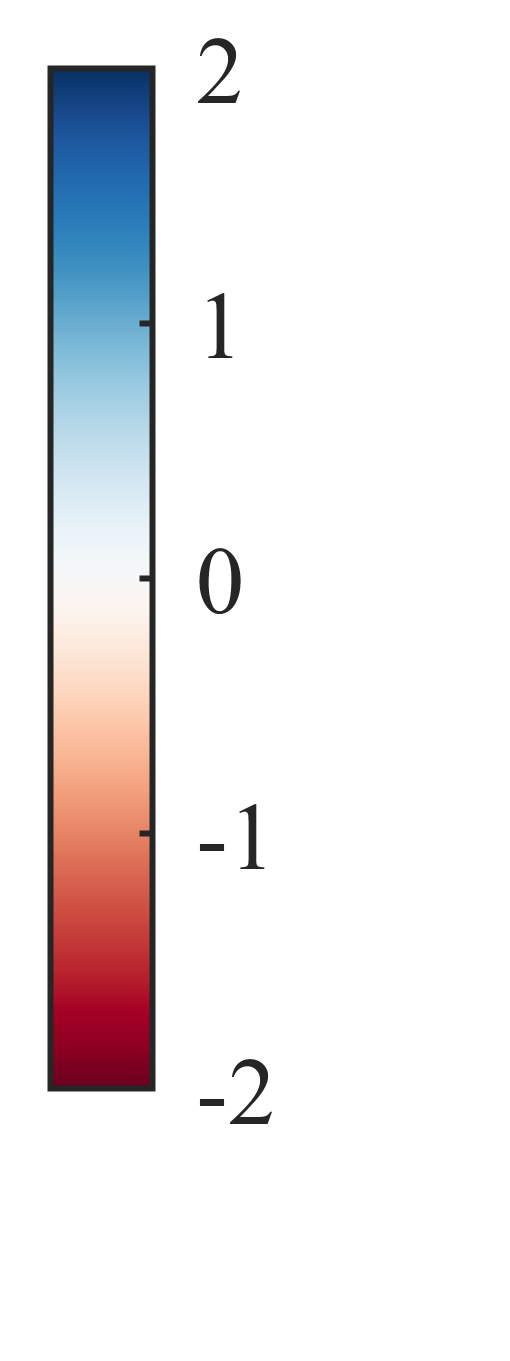}}\\
\addtocounter{subfigure}{-1}
\subfloat[$\real(E_x)$]{\begin{tabular}[b]{c}%
\includegraphics[trim={11px 9px 5px 2px},clip]{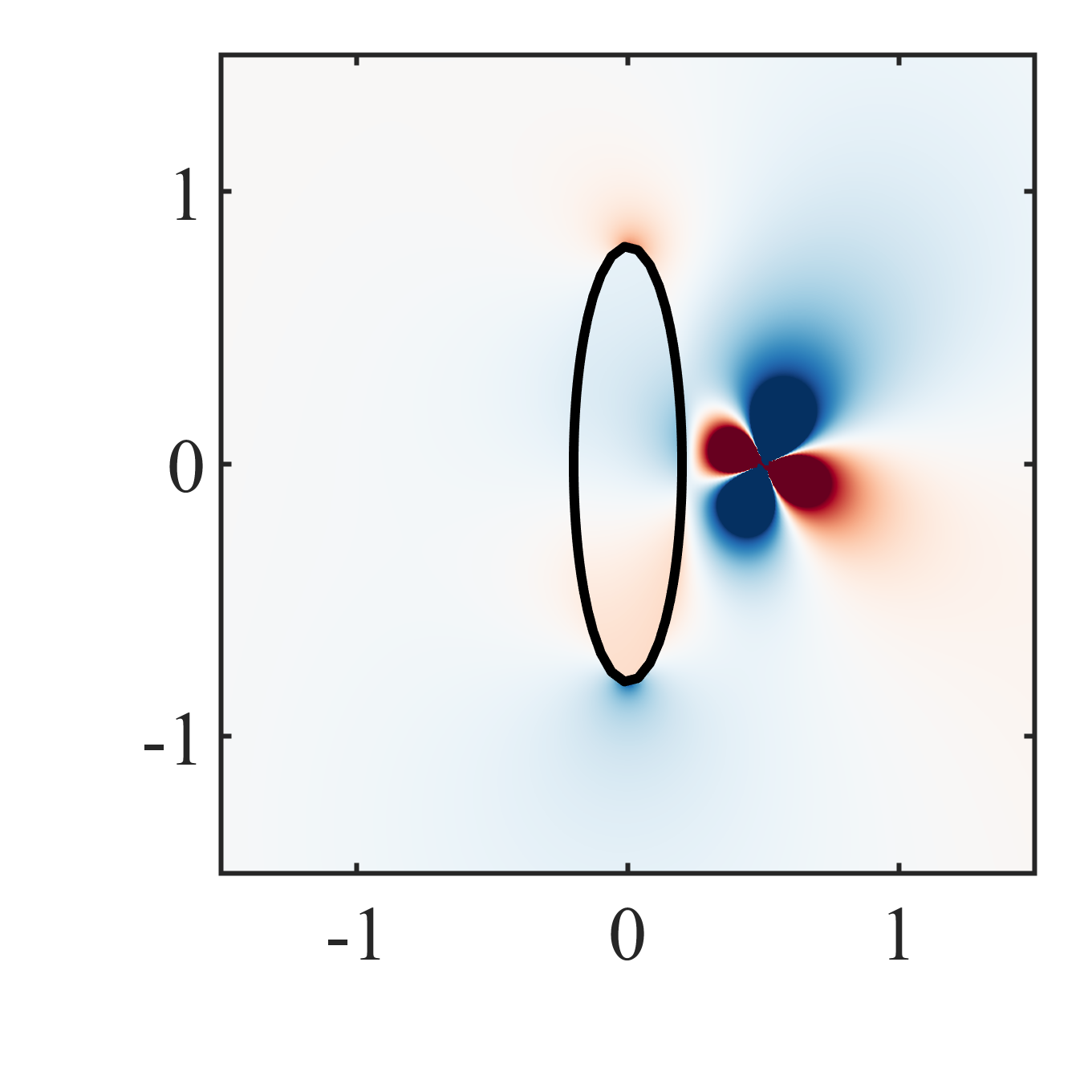}
\end{tabular}}
\subfloat[$\real(E_y)$]{\begin{tabular}[b]{c}%
\includegraphics[trim={11px 9px 5px 2px},clip]{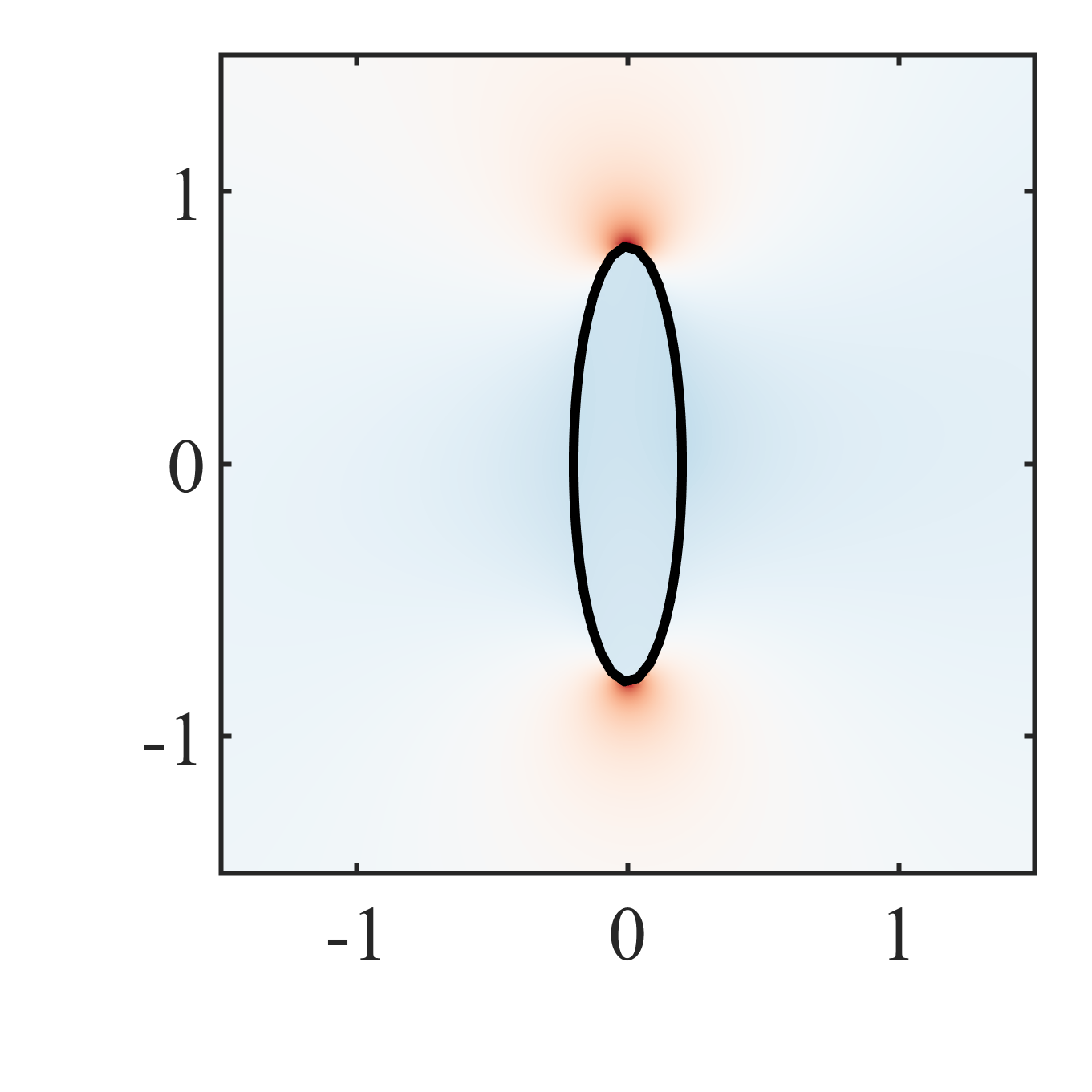}
\end{tabular}}
\subfloat{
\includegraphics[trim={3px 9px 12px 2px},clip]{dipoleaxis.png}}
\caption{The total field produced by a monochromatic dipole source at (0.5, 0) with orientation (1, 1, 0), near a metallic ellipse of permittivity $\epsilon_i = - 5.3 + 0.22i$.}
\label{fig:dipoleeps}
\end{center}
\end{figure}

We now place a near-field dipolar source at $(x, y) = (0.5, 0)$ with a diagonal orientation, $(p_x, p_y, p_z)/\epsilon_0 = (1, 1, 0)/\sqrt{2}$, and set the inclusion permittivity to $\epsilon_i = -5.3 + 0.22i$. The dipole strength is $|\bv{p}|/\epsilon_0 = 1$ Vm, where $\bv{p}$ is the dipole moment per unit length in the perpendicular direction. This choice of units allows the figures to simultaneously represent the electric field, in units of Vm$^{-1}$, and the Green's tensor, which is a unitless quantity in 2D. We use GENOME \eqref{eq:genome} to expand the total field, employing all of modes found above without any filtering. Figure \ref{fig:dipoleeps} shows the real and imaginary parts of the total $E_x$ and $E_y$ fields. Since the dipole has zero component in the $z$-direction, the $E_z$ field should be identically zero, which our simulation again reproduces to numerical accuracy. Since we did not filter any of the modes, the results are potentially polluted by higher-order modes, which are liable to be inaccurate due to truncation. However, no such ill effects were detected, demonstrating the absence of any spurious modes and the reliability of the method and the basis.

\subsection{Convergence properties}
\label{sec:convergence}
\begin{figure}[!t]
\begin{center}
\includegraphics{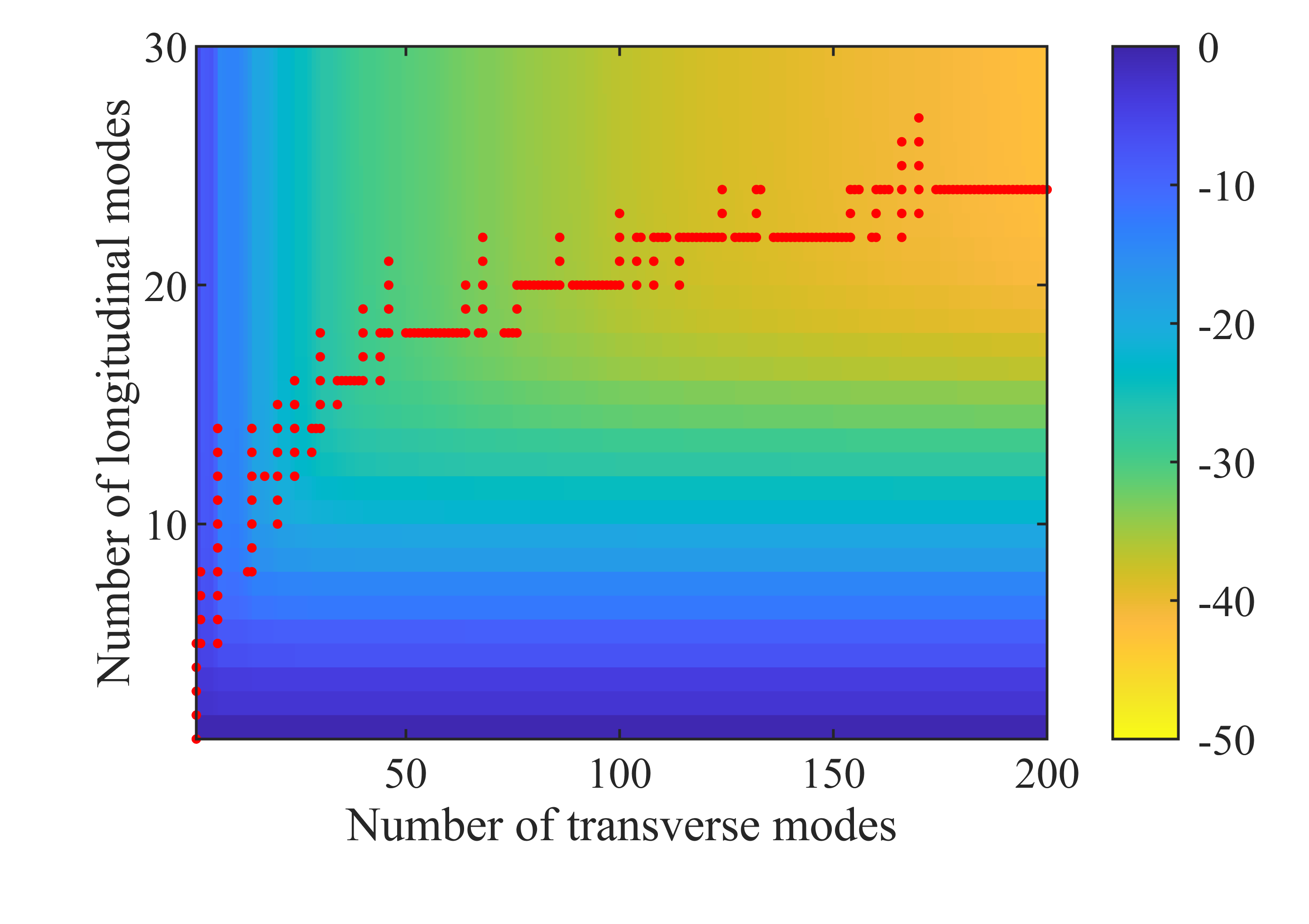}
\caption{Shows the convergence in the eigenvalue of the eigenmode displayed in Figure \ref{fig:plasmodeelli} on a decibel scale, as a function of longitudinal and transverse modes. The red dots trace the path taken by the line in Figure \ref{fig:totmode}.}
\label{fig:ltmode}
\end{center}
\end{figure}

\begin{figure}[!t]
\begin{center}
\includegraphics{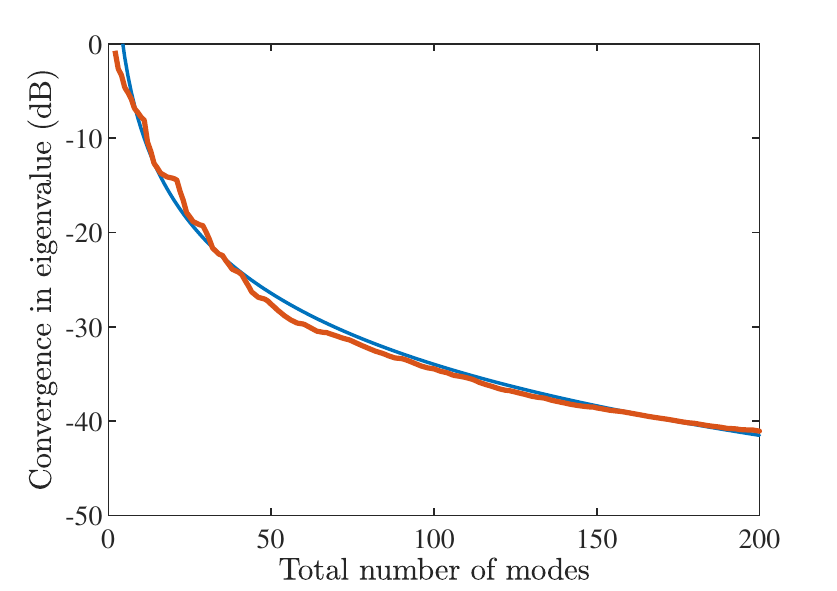}
\caption{Similar to Figure \ref{fig:ltmode}, with the red curve showing the convergence as a function of total number of modes. The combination of longitudinal and transverse modes used for each point is indicated by the red dots in Figure \ref{fig:ltmode}. The blue curve is added for comparison purposes, and is given by the function $f(N) = 40N^{-2.5}$, where $N$ is the total number of modes.}
\label{fig:totmode}
\end{center}
\end{figure}

\begin{figure}[!t]
\begin{center}
\includegraphics[trim={0px 35px 12px 45px},clip]{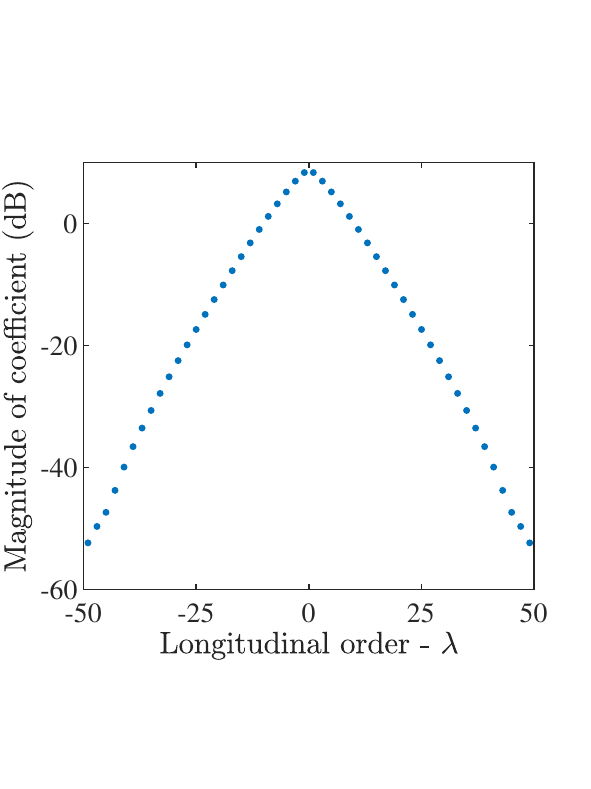}
\caption{Shows the contribution of each longitudinal basis mode during the expansion of the plasmonic mode displayed in Figure \ref{fig:plasmodeelli}. Magnitudes of the coefficients are plotted on a logarithmic scale.}
\label{fig:lmodeconv}
\end{center}
\end{figure}

\begin{figure}[!t]
\begin{center}
\includegraphics{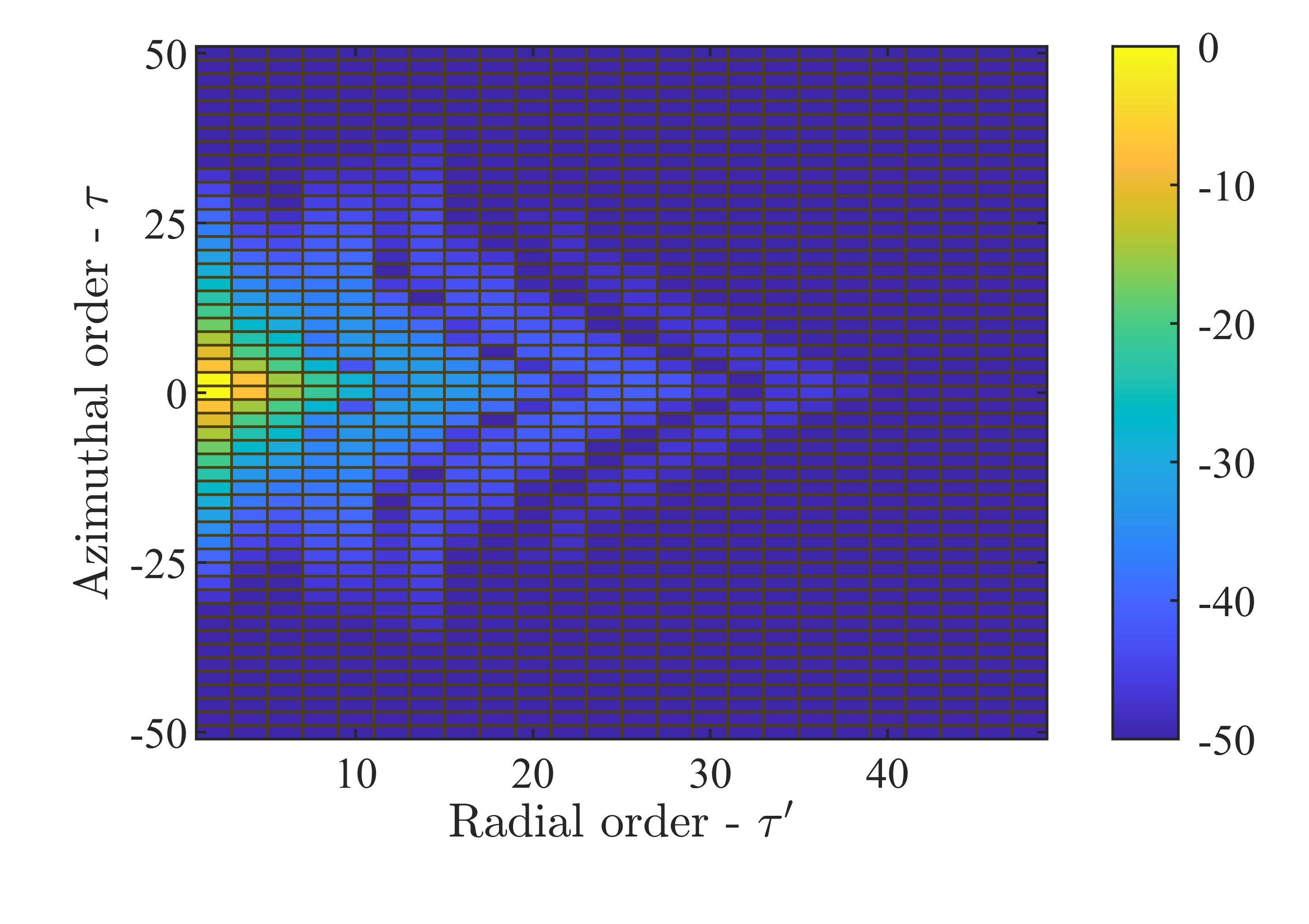}
\caption{As in Figure \ref{fig:lmodeconv} but shows the contribution of each transverse mode, arranged by azimuthal and radial order. The color scale represents the magnitude of the coefficients on a logarithmic scale.}
\label{fig:tmode}
\end{center}
\end{figure}

\begin{figure}[!t]
\begin{center}
\subfloat[]{\begin{tabular}[b]{c}%
\includegraphics[trim={0px 35px 12px 45px},clip]{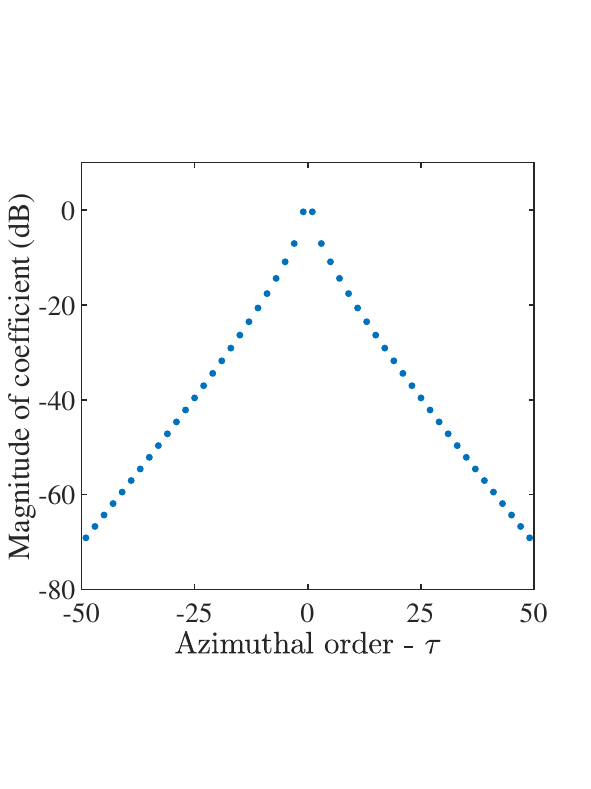}
\end{tabular}}
\subfloat[]{\begin{tabular}[b]{c}%
\includegraphics[trim={10px 35px 12px 45px},clip]{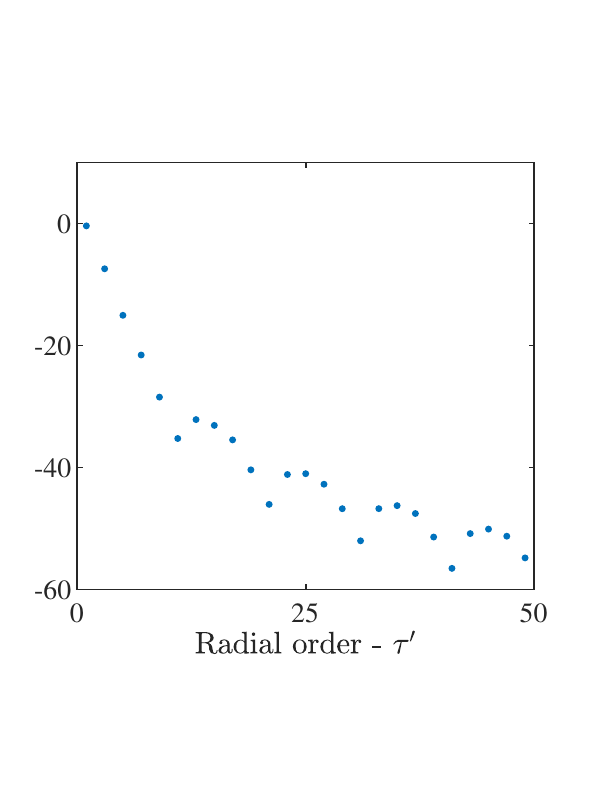}
\end{tabular}}
\caption{Duplicates the data of Figure \ref{fig:tmode}, but on two separate line plots. The magnitudes are plotted with respect to azimuthal order while the radial order is fixed to 1, then the magnitudes are plotted with respect to radial order while the azimuthal order is fixed to 1.}
\label{fig:tmodeline}
\end{center}
\end{figure}

We now analyze the convergence properties of the discontinuous longitudinal modes, and the re-expansion method in general, in generating modes of the elliptical geometry of Section \ref{sec:ellipse}. In particular, we choose to study its brightest plasmonic mode, displayed in Figure \ref{fig:plasmodeelli}. This mode is important for simulating Maxwell's equations via \eqref{eq:genome} and is also challenging to generate using other simulation methods. We shall consider two different measures of convergence.

We first demonstrate the convergence in the eigenvalue of the brightest plasmonic mode as more discontinuous longitudinal \eqref{eq:Lmodecyl} and transverse \eqref{eq:disprel} embedding modes are included in the expansion \eqref{eq:expansion}. We compare against a reference eigenvalue computed from a more accurate calculation performed in Section \ref{sec:ellipse}, plotting in Figure \ref{fig:ltmode} on a decibel scale the normalized difference $10\log_{10}|(\epsilon_m-\epsilon_\textrm{ref})/\epsilon_\textrm{ref}|$. More transverse embedding modes than longitudinal embedding modes are necessary for convergence, with $4-5$ digits of precision eventually attained with the parameters considered. It is also useful to consider the convergence with the total number of embedding modes, whether transverse or longitudinal. We thus extract from Figure \ref{fig:ltmode} the combination of transverse and longitudinal modes that gives the best agreement with $\epsilon_\textrm{ref}$, indicated by the series of red dots. Figure \ref{fig:totmode} plots the convergence along this trajectory. This shows that convergence to $10^{-4}$ can be achieved using approximately 200 modes, which, from Figure \ref{fig:ltmode} consists of approximately only 23 longitudinal modes with the remainder being transverse modes.

These plots also demonstrate that the discontinuous longitudinal modes are responsible for most of the initial convergence, especially the first 15 modes. Further convergence is only achieved by including many more transverse modes, and the eventual rate of convergence is determined by the number of transverse modes, masking the rapid convergence of the discontinuous longitudinal modes. Here, the convergence appears to be governed by a power law, and is approximately $N^{-2.5}$, where $N$ is the total number of modes. For comparison, a blue reference curve is plotted in Figure \ref{fig:totmode}, given by $f(N) = 40N^{-2.5}$. This scaling is comparable with other methods that expand the modes of an open system using the modes of a simpler open system.\autocite{doost2014resonant,lobanov2019resonant,chen2020efficient}

We now consider convergence from a different angle, to disentangle and isolate the convergence properties of the discontinuous longitudinal embedding modes from the transverse modes.  We consider the contribution of each embedding mode $\mu$ to the highly accurate reference calculation of the chosen target mode $m$, simply by plotting the magnitudes of each expansion coefficient $|c_{\mu,m}|$. The index $\mu$ is a combined index that can represent either longitudinal or transverse modes. This analysis is also useful for determining truncation limits for each type of mode. We first consider the discontinuous longitudinal modes, plotting their magnitudes in Figure \ref{fig:lmodeconv}. Exponential convergence is clearly evident. The transverse modes are indexed by azimuthal order $\tau$ and radial order $\tau'$, with their magnitudes indicated by the color scale of Figure \ref{fig:tmode}. The fundamental azimuthal and radial orders provide the greatest contribution, with the weights of higher orders progressively diminishing. We further isolate the convergence along $\tau$ and $\tau'$ in Figure \ref{fig:tmodeline}, plotting the magnitudes while holding the other parameter constant. This demonstrates the exponential decay of the the coefficients with azimuthal but not radial order.

\section{Discussion and conclusion}
In this paper, we find the modes of a target system open system \eqref{eq:eigendiff} by expanding using embedding modes of a simpler open system \eqref{eq:eigenb}, in a process we call re-expansion. We resolve the Gibbs phenomenon by including among the embedding basis a subset that is discontinuous, defined by \eqref{eq:potendef}, \eqref{eq:dirichlet}, and \eqref{eq:Lmodecyl}. This discontinuous set is constructed by incorporating knowledge that field discontinuities occur only at step interfaces, with fields otherwise being smooth. In the context of Maxwell's equations, there exists a physical interpretation for field discontinuities, corresponding to a non-zero divergence of the field. Thus, we construct a longitudinal set of modes to reproduce the discontinuity, with a simple definition that can be adapted to any curved surface. Conveniently, these longitudinal modes emerge as static-like solutions of electrodynamic Maxwell's equations \eqref{eq:eigenb}, and so are inherently complementary to the usual transverse embedding modes. The definition via a Laplace equation also means that they are easily found using standard numerical techniques.

The two key advantages of the discontinuous longitudinal embedding modes are convergence and reliability. Rapid convergence in turn implies speed and efficiency. Convergence is analyzed in Section \ref{sec:convergence}. We focus on the fundamental plasmonic mode, since our method is equally adept at finding this mode as any other mode, but such modes are difficult to find using other methods. Figure \ref{fig:lmodeconv} shows that the convergence rate is exponential with longitudinal order $\lambda$ confirming expectations set forth near the end of Section \ref{sec:longitudinal}. This efficiency is because the modes need only be defined along the interface, enabling the dimension of the basis to be reduced by one. Furthermore, the complex amplitude of the discontinuity varies smoothly along the interface, which Fourier series can capture with exponential convergence. The contribution of the discontinuous longitudinal embedding modes to the target mode is large, so the overall convergence rate of the re-expansion method is initially rapid, as seen in Figures \ref{fig:ltmode} and \ref{fig:totmode}.

These convergence properties contrast with that of the transverse embedding basis, which is smooth in its interior. As shown in Figures \ref{fig:tmode} and \ref{fig:tmodeline}, convergence is exponential with azimuthal order $\tau$, but is roughly polynomial with radial order $\tau'$. Convergence is slower also because there is now a 2D set of modes. This mirrors observations associated with analogous methods where modes of a target open system were expanded using the modes of an open embedding system.\autocite{chen2020efficient, lobanov2019resonant} The slower convergence properties of the transverse modes ultimately limits the convergence rate of the re-expansion method, as seen is Figure \ref{fig:totmode}.

Another advantage of the discontinuous longitudinal modes \eqref{eq:Lmodecyl}, perhaps even more important than speed, is the reliability they impart. Since it is far more constrained than the Fourier-Bessel basis \eqref{eq:FBmode}, it has limited freedom for generating numerical noise, and prevents spurious modes from appearing. There are two spurious modes that typically afflict other modal solvers. Firstly, for methods that discretize space such as the finite element method, plasmonic behavior is liable to produce artificially localized modes. This occurs with triangular or tetrahedral mesh elements, creating sharp points along the interface during discretization that cause field divergence during solution. Only a very fine discretization can prevent their appearance. However, such spurious modes are never generated using the discontinuous longitudinal modes \eqref{eq:Lmodecyl}, since this basis is smooth along the interface and contains no singularities. For practical purposes though, this ultimate limit is less relevant than the rapid initial convergence to 2-3 digits.

Secondly, an infinite number of longitudinal modes of the type \eqref{eq:potendef} exist. As explained in Section \ref{sec:longitudinal}, their modal fields can be arbitrary. While the longitudinal modes of the embedding geometry are useful for defining a discontinuous basis for expanding the modes of target geometry, the longitudinal modes of the target geometry are largely useless in subsequently expanding Maxwell's equations via GENOME \eqref{eq:genome}. In this sense, the longitudinal modes of the target geometry can be considered spurious. Methods that do not enforce transversality can generate many such modes, whose modal fields are typically meaningless when found numerically. Ordinarily, such spurious modes can be identified and discarded based on their eigenpermittivity $\epsilon_m$, which should be zero. But some spurious longitudinal modes are generated with non-zero $\epsilon_m$, possibly due to numerical noise, and these may require visual inspection to identify. Even the mere presence of spurious longitudinal target modes can be hazardous, particularly for iterative eigenvalue solvers, as there exists an infinite number of such modes that serve as attractors. Such spurious target modes also cannot be generated using the discontinuous longitudinal embedding basis, since longitudinal component of the latter has been constrained to the interface of the target geometry, eliminating the freedom to generate arbitrary longitudinal target modes.

Overall, use of the discontinuous longitudinal embedding modes enabled the re-expansion method \eqref{eq:perteig} to generate many accurate modes of the target geometry. Exponential convergence means that each discontinuous longitudinal mode \eqref{eq:Lmodecyl} is capable of replacing many hundreds of Fourier-Bessel longitudinal modes \eqref{eq:FBmode}. This drastically reduces the number of modes needed to solve \eqref{eq:perteig}, expediting simulation times. With very few embedding modes, the target modes were quickly obtained to approximately 3 digits of accuracy. Further accuracy is also possible, but hinged instead on the slower convergence behavior of the transverse embedding modes. A complete set of target modes is obtained, up to truncation, which is important since missing modes negatively impact the subsequent expansion of Maxwell's equations via GENOME \eqref{eq:genome}. Many plasmonic modes are also found, which is often difficult to achieve using other methods due to their tight field confinement. Finally, no spurious modes of the target geometry are produced. Spurious modes pollute the expansion \eqref{eq:genome}, also leading to erroneous results. Thus, manual inspection or filtering is unnecessary prior to use. This reliability means the method is suitable for automation and scripting if desired.

Finally, we have focused on piecewise uniform permittivity profiles consisting of a step interface whose locus can be traced as a smooth function in polar form \eqref{eq:Lmodecyl}. However, the analysis of Section \ref{sec:longitudinal} is flexible and future work will consider extensions to more complex step interfaces featuring corners and junctions. Future work may also consider permittivity profiles with both smooth variations and step jumps.

\section{Acknowledgments}
The authors would like to acknowledge Egor A.\ Muljarov for many useful discussions.
\printbibliography
\typeout{get arXiv to do 4 passes: Label(s) may have changed. Rerun}
\end{document}